\documentclass[showpacs,showkeys,preprintnumbers,pre,amsmath,amssymb,superscriptaddress]{revtex4}

\usepackage{amssymb}
\usepackage{amsmath}
\usepackage{graphicx}
\usepackage{color}
\usepackage{hyperref}
\usepackage{subfigure}
\usepackage{amssymb}
\usepackage{enumerate}
\usepackage{amsthm,amsfonts,amsmath}
\usepackage{bm}
\usepackage{graphicx}
\usepackage{graphics,epsf,psfrag,pifont,dsfont,bbold}
\usepackage{siunitx}
\usepackage{xspace}
\usepackage{accents}

\newcommand{\be}{\begin{equation}}
\newcommand{\ee}{\end{equation}}
\newcommand{\ben}{\begin{equation*}}
\newcommand{\een}{\end{equation*}}
\newcommand{\vc}{v_{\rm c}}
\newcommand{\vd}{v_{\rm d}}
\newcommand{\vm}{v_{\rm m}}
\newcommand{\vs}{v_{\rm s}}
\newcommand{\Qc}{Q_{\rm c}}
\newcommand{\Qd}{Q_{\rm d}}
\newcommand{\etac}{\eta_{\rm c}}
\newcommand{\etad}{\eta_{\rm d}}
\newcommand{\kappac}{\kappa_{\rm c}}
\newcommand{\kappad}{\kappa_{\rm d}}
\newcommand{\GT}{G_{T}}
\newcommand{\tshear}{\tau_{\mbox{\tiny {shear}}}}

\newcommand{\Ca}{\mbox{Ca}}
\newcommand{\Ma}{\mbox{Ma}}

\newcommand{\Cam}{\mbox{Ca}_{\mbox{\tiny{m}}}}
\newcommand{\Ren}{\mbox{Re}}

\DeclareMathOperator{\sgn}{sgn}

\newcommand{\delF}[1]{\textcolor{blue}{}}

\definecolor{dBgreen}{rgb}{.3,.5,.45}

\begin{document}

\title{Lattice Boltzmann Simulations of Droplet formation in confined Channels with Thermocapillary flows}

\author{A. Gupta}
\email{anupam1509@gmail.com}
\affiliation{Department of Physics \& INFN, University of Rome ``Tor Vergata'', Via della Ricerca Scientifica 1, 00133, Rome, Italy.}

\author{M. Sbragaglia}
\email{sbragaglia@roma2.infn.it}
\affiliation{Department of Physics \& INFN, University of Rome ``Tor Vergata'', Via della Ricerca Scientifica 1, 00133, Rome, Italy.}

\author{D. Belardinelli}
\email{belardinelli@roma2.infn.it}
\affiliation{Department of Physics \& INFN, University of Rome ``Tor Vergata'', Via della Ricerca Scientifica 1, 00133, Rome, Italy.}

\author{K. Sugiyama}
\email{kazuyasu.sugiyama@me.es.osaka-u.ac.jp}
\affiliation{Department of Mechanical Science and Bioengineering, Graduate School of Engineering Science, Osaka University 1-3 Machikaneyama, Toyonaka, Osaka 560-8531, Japan}

\date{\today}

\begin{abstract}
Based on mesoscale lattice Boltzmann simulations with the ``Shan-Chen'' model, we explore the influence of thermocapillarity on the break-up properties of fluid threads in a microfluidic T-junction, where a dispersed phase is injected perpendicularly into a main channel containing a continuous phase, and the latter induces periodic break-up of droplets due to the cross-flowing. Temperature effects are investigated by switching on/off both positive/negative temperature gradients along the main channel direction, thus promoting a different thread dynamics with anticipated/delayed break-up. Numerical simulations are performed at changing the flow-rates of both the continuous and dispersed phases, as well as the relative importance of viscous forces, surface tension forces and thermocapillary stresses. The range of parameters is broad enough to characterize the effects of thermocapillarity on different mechanisms of break-up in the confined T-junction, including the so-called ``squeezing'' and ``dripping'' regimes, previously identified in the literature. Some simple scaling arguments are proposed to rationalize the observed behaviour, and to provide quantitative guidelines on how to predict the droplet size after break-up. 
\end{abstract}

\maketitle

\section{Introduction}

Droplet control and production in microfluidic devices has gained a considerable importance, due to the ample variety of applications involved. Droplets can indeed transport very small volumes of fluids, where reagents can be stored and analyzed for a variety of microfluidic applications~\cite{Christopher07,Seeman12,Christopher08,Teh08,Baroud10,Thorsen01,Glawdeletal,Glawdeletalb,Glawdeletalbb,Wheeler08,Garstecki06,Gong08,Baroud07,Delville12,selva2011temperature,Yap09,Zhangetal12,Zhangetal13,Zhangetal14}. In this context, droplet manipulation based on thermocapillary stresses has also been established~\cite{Nguyen06,Baroud07,Yap09,Murshed09,Ho11,Delville12,Zhangetal12,Zhangetal13,Zhangetal14,Li15}. Thermocapillary stresses arise because of a surface tension gradient induced by temperature differences, with the surface tension at the interface between two fluids being a decreasing function of the temperature in most of the practical situations encountered~\cite{Baroud07,Nguyen07,Yap09,Ho11,Delville12}. The resulting ``Marangoni flows'' induce droplet motion from the regions of high surface tension to the regions of low surface tension. The focus of this paper is on mesoscopic numerical simulations to quantify the effects of thermocapillary forces on geometry mediated droplet production in confined channels. For the confined channel, we use a T-junction geometry~\cite{Demenech07,Demenech06}, where a dispersed (d) phase is injected from a side channel perpendicularly into a main channel containing a continuous (c) phase. The (periodic) production of droplets is then induced by cross-flowing and is affected by the Capillary number $\Ca$ - which quantifies the importance of the viscous forces with respect to the surface tension forces - and by the flow-rate ratio $\phi=\Qd/\Qc$ of the two immiscible fluids~\cite{Demenech07,Garstecki06,LiuZhang09,BowerLee11,gupta2016effects,gupta2016lattice}. Distinct mechanisms of droplet formation have been identified~\cite{Demenech07,Garstecki06}: these include both the so-called ``squeezing'' and ``dripping'' regimes~\cite{Demenech07,Demenech06,LiuZhang11,Garstecki06}. Specifically, in the squeezing mechanism the break-up process is determined by the build-up of pressure in the main channel caused by the obstruction of the dispersed phase entering from the side channel~\cite{Demenech07,Garstecki06}. In the dripping regime, instead, shear forces start do be relevant and promote a smaller droplet size if compared to the squeezing regime~\cite{Demenech07,Garstecki06}. The Newtonian problem has been very well characterized in a series of papers in the literature, both experimentally and numerically (see~\cite{Demenech07,Demenech06,LiuZhang11,Garstecki06,LiuZhang09,Liu15} and references therein). In this paper, we aim to take a step forward, by exploring the influence of thermocapillarity on the dynamics and break-up properties of fluid threads in the confined T-junction geometry. Our numerical approach is based on lattice Boltzmann (LB) models~\cite{Benzi92,Succi01,Zhang11,Aidun10}, which proved to be very flexible numerical tools for the simulation of a wide variety of problems involving droplets dynamics in microfluidic geometries~\cite{Xi99,vandersman08,Komrakova13,Liuetal12,Liu15,Moradi,Thampi,Liuetal12,Gupta09,Gupta10,BowerLee11,gupta2016effects,gupta2016lattice}. LB allows to solve the diffuse-interface hydrodynamic equations of a binary mixture of two components~\cite{Yue04,Yueetal05,Yueetal06a,Yueetal06b,Yueetal08,Yueetal12} in presence of phase segregation, which confines ``pure'' fluids to different regions, separated by a non ideal interface with a positive surface tension. Multiphase flows with thermocapillary effects have already been investigated in the context of the LB methodology. In particular, Liu {\it et al.}~\cite{Zhangetal12,Zhangetal13,Zhangetal14,Li15,LiuZhang15,Liu17} developed LB multiphase models to account for thermocapillary flows. The LB model proposed by Liu {\it et al.}~\cite{Zhangetal12,Zhangetal13,Zhangetal14,Li15} relies on an ``improved color-fluid'' LB scheme~\cite{Gunstensen91}, in which the non-ideal forces and the thermocapillary stresses are introduced using the idea of ``continuum surface force''~\cite{Zhangetal12,Zhangetal13,Zhangetal14,Li15,LiuZhang15}. The authors have successfully described droplets dynamics and migration, and the effects of localized heating sources in confined environments. Our study differs from the works by Liu {\it et al.}~\cite{Zhangetal12,Zhangetal13,Zhangetal14,Li15,LiuZhang15,Liu17} both in terms of the LB methodology and in terms of the physical problem considered. From the methodological side, the corresponding modelling for the thermocapillary flows is developed on the top of the ``Shan-Chen'' model for non-ideal binary mixtures~\cite{SC93,SC94,Shan08,CHEM09,SbragagliaBelardinelli}. This may be useful, for example, in those situations where structured fluids and emulsion droplets~\cite{CHEM09,Sbragaglia12} need to be simulated.\\
The paper is organized as follows. In section~\ref{sec:numerics} we will present the necessary mathematical background for the problem studied, showing the equations that we consider in both the continuous and dispersed phases. In subsections~\ref{sec:LBmulti}-\ref{sec:LBTHERMAL} we will recall the basic details of the LB scheme that we use to solve the fluid dynamical equations. In subsection~\ref{sec:thermo} we will illustrate the construction of the thermocapillary coupling on the top of the ``Shan-Chen'' model. In section~\ref{sec:TJsetup} we will apply the numerical methodology to the confined T-junction geometry, and we will provide quantitative details on how the thermocapillary effects influence the transition from the squeezing dominated regime to the dripping dominated regime. We will finally rationalize the observed behaviour and provide quantitative guidelines on how to predict the droplet size after break-up. Conclusions will follow in section~\ref{sec:conclusions}. Useful benchmark tests and validation studies will be discussed in Appendix~\ref{sec:benchmark}. Some technical details on the problem of the motion of a cylindrical droplet subject to a uniform temperature gradient are reported in Appendix~\ref{sec:Appendix}.

\section{Continuum Equations and Numerical method}\label{sec:numerics}

The equations we consider in the continuous and dispersed phase are the Navier-Stokes (NS) equations in the form (subscripts c/d refer to the continuous/dispersed phase)
\be\label{NScd}
\begin{split}
\rho_{i} \left[ \partial_t \bm u_{i} + ({\bm u}_{i} \cdot {\bm \nabla}) \bm u_{i} \right]  =  - {\bm \nabla}P_{{\rm b},i}+ {\bm \nabla} \cdot \left(\eta_{i} ({\bm \nabla} {\bm u}_{i}+({\bm \nabla} {\bm u}_{i})^{T})\right) \hspace{.2in} i={\rm c,d}.
\end{split}
\ee
Here, ${\bm u}$ and $\eta$ are the velocity and the dynamic viscosity, respectively. $\rho$ is the total density, $P_{\rm b}$ the bulk pressure, and $({\bm \nabla} {\bm u})^T$ the transpose of $({\bm \nabla} {\bm u})$. Immiscibility between the dispersed phase and the continuous phase is introduced using the so-called ``Shan-Chen'' model~\cite{SbragagliaGuptaScagliarini,SC93,SC94,Shan08}, which ensures phase separation with the formation of stable (diffuse) interfaces between the two phases with a positive surface tension $\sigma$. An additional interface force is introduced due to the temperature-dependent surface tension $\sigma=\sigma(T)$. The associated interfacial force is most conveniently written in the limit of a sharp interface, using the differential operator ${\bm \nabla}_{S}=({\bm \delta}-{\hat {\bm n}} \, {\hat {\bm n}}) \cdot {\bm \nabla}$, providing projection tangentially to the interface~\cite{Zhangetal12,Zhangetal13,Zhangetal14,Li15} 
\begin{equation}\label{eq:additionalcoupling}
{\bf F}={\bm \nabla} \cdot (\sigma(T) ({\bm \delta}-{\hat {\bm n}} \, {\hat {\bm n}}) \delta_{{\cal S}})=\sigma \, {\cal C} \, \delta_{{\cal S}} \, {\bm n}+ ({\bm \nabla}_{S} \sigma) \delta_{{\cal S}}
\end{equation}
where $\delta_{{\cal S}}$ is the Dirac delta function and ${\cal C}$ is the (local) curvature at the interface~\cite{Zhangetal12,Zhangetal13,Zhangetal14,Li15}. Equivalent expressions of~\eqref{eq:additionalcoupling} for the general case of diffuse interfaces are provided in subsection~\ref{sec:thermo}. In presence of thermocapillarity, we need to relate the surface tension to the temperature. For the sake of simplicity, we only consider a linear relation between the surface tension and the temperature $\sigma(T)=\sigma_{0}+\sigma_T (T-T_{0})$, where $T_{0}$ is a reference temperature, $\sigma_{0}$ is the surface tension at $T_{0}$, $\sigma_T=d\sigma/dT$ is the rate of change of surface tension with temperature and is an input parameter in our numerical simulations. Together with the macroscopic NS equations~\eqref{NScd}, we have considered the following advection-diffusion equation (i.e. passive-scalar equation) for the temperature field in both the continuous and the dispersed phase~\cite{Ripesi14}
\begin{equation}\label{THERMAL}
\partial_t T_i + ({\bm u}_{i} \cdot {\bm \nabla}) T_i={\bm \nabla} \cdot (\kappa^{(T)}_{i} {\bm \nabla} T_i) \hspace{.2in} i={\rm c,d}
\end{equation}
where $\kappa^{(T)}$ is the thermal diffusivity. Equations~\eqref{NScd}-\eqref{THERMAL} are solved by means of an algorithm combining a multicomponent LB model coupled to an LB solver for the advection-diffusion equation~\eqref{THERMAL}. Technical details on the multicomponent LB and the advection-diffusion LB scheme are not new, hence they are just briefly recalled in subsections \ref{sec:LBmulti} and \ref{sec:LBTHERMAL}. In subsection \ref{sec:thermo}, instead, we will discuss more extensively on the modelling of thermocapillary effects, since it constitutes a novelty from the methodological point of view for the paper.

\subsection{Modelling of Multicomponent Fluid}\label{sec:LBmulti}

The multicomponent LB takes into account the evolution dynamics of the discretized probability density function $f_{\ell s}({\bm{x}},t)$ to find at position ${\bm{x}}$ and time $t$ a fluid particle of component $\ell=A,B$ with velocity  ${\bm{c}}_{s}$. The numerical simulations described in the paper make use of the D3Q19 model \footnote{For the 2D simulations we shrink one dimension} whose discrete velocity set is given by
\begin{equation}\label{velo}
{\bm{c}}_{s}=
\begin{cases}
(0,0,0) & s=0\\
(\pm 1,0,0), (0,0,\pm 1) , (0,\pm 1,0), & s=1\ldots6\\
(\pm 1,\pm 1,0), (0,\pm 1,\pm 1), (\pm 1,0,\pm 1) & s=7\ldots18
\end{cases}.
\end{equation}
A majority of one of the two components then identifies the dispersed and the continuous phases in the hydrodynamic equations~\eqref{NScd}. The LB dynamics accounts (over a unitary time step) for both the relaxation and the streaming of $f_{\ell s}({\bm{x}},t)$~\cite{Succi01}:
\begin{equation}\label{EQ:LBapp}
f_{\ell s} ({\bm{x}} + {\bm{c}}_{s} , t + 1)-f_{\ell s} ({\bm{x}}, t) = \sum_{j} {\cal A}_{s j}(f_{\ell j}-f^{(eq)}_{\ell j}) + {\cal S}^{g}_{\ell s}
\end{equation}
where the matrix terms ${\cal A}_{s j}$ (the same for both components) drive the relaxation towards the local equilibrium~\cite{Succi01} 
\be\label{feq}
f_{\ell s}^{(eq)}=w_{s} \rho_{\ell} \left[1+\frac{{\bm{u}} \cdot {\bm{c}}_{s}}{c_S^2}+\frac{{\bm{u}}{\bm{u}}:({\bm{c}}_{s}{\bm{c}}_{s}-{\bm {\delta}})}{2 c_S^4} \right]
\ee
where $c_S=1/\sqrt{3}$ is the isothermal speed of sound, ${\bm{u}}$ is the fluid velocity, and the weights $w_{s}$ are given by $w_{s}=1/3$ for $s=0$, $w_s=1/18$ for $s=1 \ldots 6$ and $w_s=1/36$ for $s=7 \ldots 18$. We work in the framework of the multiple relaxation time (MRT) models~\cite{Dunweg07,DHumieres02,Premnath,SegaSbragaglia13}. This offers the particular advantage to control the various processes (mass diffusion, momentum diffusion, etc) independently. The MRT models make use of a ``mode space'' identified with a suitable basis vectors ${\bm{b}}_{k}$ ($k=0,...,18$), hence the ``modes'' calculated as $M_{\ell k}=\sum_{s} {\bm{b}}_{k s} f_{\ell s}$ ($k=0,...,18$)~\cite{Dunweg07,DHumieres02,Premnath,SegaSbragaglia13}. The collisional term then describes a linear relaxation of the modes, $M_{\ell k} \rightarrow (1+a_k) M_{\ell k}+M_{\ell k}^{g}$, where the relaxation frequencies $-a_k$ are related to the transport coefficients in the hydrodynamic equations (see also below). The lowest order modes are directly linked to the hydrodynamic variables as follows. The density of both components ($\rho_{\ell}$) and the total density ($\rho$) coincide with the zero-th order modes
$$
\rho_{\ell}=M_{\ell 0}=\sum_{s} f_{\ell s} \hspace{.2in} \rho=\sum_{\ell}M_{\ell 0} =\sum_{\ell}\rho_{\ell}
$$ 
while the next three modes $(M_{\ell 1}, M_{\ell 2}, M_{\ell 3})$ define the velocity of the mixture 
\be\label{totmom}
{\bm{u}}=(u_x,u_y,u_z) \equiv \frac{1}{\rho}\sum_{\ell} (M_{\ell 1}, M_{\ell 2}, M_{\ell 3})   +\frac{(g_x,g_y,g_z)}{2 \rho} = \frac{1}{\rho}\sum_{\ell} \sum_{s} (f_{\ell s} {\bm c}_{sx},f_{\ell s} {\bm c}_{sy},f_{\ell s} {\bm c}_{sz})+\frac{(g_x,g_y,g_z)}{2 \rho}.
\ee
The higher order modes refer to the viscous stress tensor, and also other modes  which do not show-up in the hydrodynamic limit~\cite{Dunweg07,DHumieres02,Premnath,SegaSbragaglia13}. The term ${\cal S}_{\ell s}^{g}$ is a forcing source embedding the effects of a forcing term with density ${\bm{g}}_{\ell}$~\cite{Dunweg07,Premnath,SegaSbragaglia13}. The term ${\bm{g}}=\sum_{\ell} {\bm g}_{\ell}$ in Eq.~\eqref{totmom} is the total force. Given the relaxation frequencies of the momentum ($-a_{\rm M}$), bulk ($-a_{\rm b}$) and shear ($-a_{\rm s}$) modes, the forcing source that ensures the correct recovery of the hydrodynamic equations can be computed exactly~\cite{SegaSbragaglia13}. The hydrodynamic evolution is obtained by considering the long-wavelength limit~\cite{Gladrow00,Succi01} of the equations for the coarse-grained density and momentum. One can indeed show that the LB scheme reproduces the continuity equations and the NS equations~\cite{Dunweg07,Premnath,SegaSbragaglia13}:
\be\label{eq:2}
\partial_t \rho_{\ell}+ {\bm \nabla} \cdot (\rho_{\ell} {\bm u}) = {\bm \nabla} \cdot {\bm D}_{\ell},
\ee
\be\label{eq:3}
\begin{split}
\rho \left[ \partial_t \bm u + ({\bm u} \cdot {\bm \nabla}) \bm u \right]= -{\bm \nabla}p+ {\bm \nabla} \cdot \left[ \eta_{\rm s} \left( {\bm \nabla} {\bm u}+({\bm \nabla} {\bm u})^{T}-\frac{2}{3} {\bm \delta} ({\bm \nabla} \cdot {\bm u}) \right) +\eta_{\rm b} {\bm \delta} ({\bm \nabla} \cdot {\bm u}) \right] + {\bm g}.
\end{split}
\ee
The viscosity coefficients are determined by the relaxation frequencies of the bulk and shear modes in~\eqref{EQ:LBapp} as $\eta_{\rm s}=-\rho c_S^2 \left(\frac{1}{a_{\rm s}}+\frac{1}{2} \right)$ and $\eta_{\rm b}=-\frac{2}{3}\rho c_S^2  \left(\frac{1}{a_{\rm b}}+\frac{1}{2} \right)$. In equations \eqref{eq:3}, $p=\sum_{\ell} p_{\ell}=\sum_{\ell} c_S^2 \rho_{\ell}$ represents the internal (ideal) pressure of the mixture, while the quantity ${\bm D}_{\ell}$ is the diffusion flux of one component into the other (see~\cite{SegaSbragaglia13} for detailed expressions). Regarding the internal forces, the ``Shan-Chen'' interaction model~\cite{SC93,SC94,Shan08,SegaSbragaglia13} for multicomponent LB schemes is used, where the force experienced by the particles of the $\ell$-th species at ${\bm x}$, is due to the particles of the other species at the neighbouring locations compatible with the lattice links
\begin{equation}\label{eq:SCforce}
{\bm g}_{\ell}({\bm{x}}) =  - g_{AB} \rho_{\ell}({\bm{x}}) \sum_{s} \sum_{\ell'\neq \ell} w_{s} \rho_{\ell^{\prime}} ({\bm{x}}+\bm{c}_{s}) {\bm c}_{s} \hspace{.2in} \ell=A,B
\end{equation}
where ${g_{AB}}$ is a parameter that regulates the intensity of phase segregating interactions. For simplicity, the sum in equation~\eqref{eq:SCforce} extends over a set of interaction links $\bm{c}_{s}$ coinciding with those of the LB dynamic~\eqref{EQ:LBapp}. For the purposes of this paper, the effect of the internal forces is more conveniently analyzed with the help of the interaction pressure tensor ${\bm P}^{(\mbox{\tiny{int}})}$~\cite{SbragagliaBelardinelli}, whose gradient is minus the internal forces, i.e., ${\bm g}=-{\bm \nabla} \cdot {\bm P}^{(\mbox{\tiny{int}})}$. The interaction pressure tensor modifies the total pressure tensor of the model, i.e., 
\begin{equation}
{\bm P}^{(\mbox{\tiny{TOT}})} = p \, {\bm \delta}+{\bm P}^{(\mbox{\tiny{int}})}.
\end{equation}
An explicit expression for ${\bm P}^{(\mbox{\tiny{int}})}$ can be derived directly on the lattice~\cite{Shan08,SbragagliaBelardinelli}. For the case of multicomponent fluids, it reads~\cite{SbragagliaBelardinelli}
\be\label{PT}
{\bm P}^{(\mbox{\tiny{int}})}({\bm x}) =\frac{1}{2} {g_{AB}} \rho_{A}({\bm x})\sum_{s} w_{s} \rho_{B}({\bm x}+{\bm c}_{s}) {\bm c}_{s} {\bm c}_{s}+\frac{1}{2} {g_{AB}} \rho_{B}({\bm x})\sum_{s} w_{s} \rho_{A}({\bm x}+{\bm c}_{s}){\bm c}_{s} {\bm c}_{s}.
\ee
Upon Taylor expanding the expression~\eqref{PT}, we get (we drop the ${\bm x}$-dependency for simplicity)
\be\label{PTexpanded}
{\bm P}^{(\mbox{\tiny{TOT}})} = \left(p+c_S^2{g_{AB}}\rho_A \rho_B + \frac{1}{4}c_S^4{g_{AB}}\rho_A \Delta \rho_B+ \frac{1}{4}c_S^4{g_{AB}}\rho_B \Delta \rho_A  \right) {\bm \delta}+\frac{1}{2} c_S^4{g_{AB}}\rho_A {\bm \nabla}{\bm \nabla}\rho_B+\frac{1}{2} c_S^4{g_{AB}}\rho_B {\bm \nabla}{\bm \nabla}\rho_A+{\cal O}({\nabla}^4)
\ee
where we recognize a bulk pressure contribution, $P_{\rm b}=p+c_S^2{g_{AB}}\rho_A \rho_B$, and other contributions which are proportional to the derivatives of both densities. These gradient terms establish a diffuse interface with a positive surface tension (see also section~\ref{sec:thermo}) whenever phase separation occurs in the model~\cite{CHEM09}. Wetting properties are introduced in the model by imposing specific values of the density in contact with the wall as described in~\cite{Benzi06,sbragaglia08}. The numerical simulations presented are carried out with ${g}_{AB}=1.5$ lbu (lattice Boltzmann units, henceforth) in~\eqref{eq:SCforce}, corresponding to a surface tension $\sigma=0.08$ lbu and associated bulk densities $\rho_A=2.1$ lbu and $\rho_B=0.12$ lbu in the $A$-rich phase. The relaxation frequencies are set to $a_{\rm s}=a_{\rm b}=-1.0$ lbu, thus reproducing the viscous stress tensor given in Eqs.~\eqref{NScd} with $\eta_{\rm c,d}=0.28$ lbu. 

\subsection{Modelling of Thermocapillary Effects}\label{sec:thermo}

In order to model the thermocapillary effects with a temperature-dependent surface tension, suitable terms need to be introduced on the top the bare pressure tensor of the model \eqref{PT}.  To this aim, we focus the attention on the mechanical balance equations at a curved two-dimensional (for simplicity) interface. In what follows, the local radius of curvature will be denoted with $R$. First of all, we notice that there exists a ``gauge invariance'' in the definition of the pressure tensor \cite{Sbragagliagauge}, in that we can define ${\bm P}^{(\mbox{\tiny{int}})}$ modulo a tensor with zero divergence. Specifically, if we introduce the pressure tensor
\be\label{eq:zeroPdiv}
\begin{split}
{\bm P}^{(0)}= & g_{AB} \left[-c_S^4  {\bm \nabla} \rho_A \cdot {\bm \nabla} \rho_B- \frac{1}{2}c_S^4 \rho_B {\Delta} \rho_A - \frac{1}{2} c_S^4 \rho_A {\Delta} \rho_B  \right] {\bm \delta}+ \\
& \frac{1}{2} g_{AB}  c_S^4 \rho_A {\bm \nabla} {\bm \nabla} \rho_B+ \frac{1}{2} g_{AB} c_S^4 \rho_B {\bm \nabla} {\bm \nabla} \rho_A+ \frac{1}{2} g_{AB} c_S^4 {\bm \nabla} \rho_A {\bm \nabla} \rho_B+ \frac{1}{2} g_{AB} c_S^4 {\bm \nabla} \rho_B {\bm \nabla} \rho_A
\end{split}
\ee
one can easily verify that \eqref{eq:zeroPdiv} has zero divergence, meaning that the pressure tensor 
$$
\tilde{\bm P}^{(\mbox{\tiny{int}})} =   {\bm P}^{(\mbox{\tiny{int}})}-{\bm P}^{(0)}
$$ 
brings the same physics of the pressure tensor ${\bm P}^{(\mbox{\tiny{int}})}$. For our purposes, we then use the pressure tensor $\tilde{\bm P}^{(\mbox{\tiny{TOT}})}={\bm P}^{(\mbox{\tiny{TOT}})}-{\bm P}^{(0)}$. Some straightforward algebraic calculations reveal the following relation between $\tilde{\bm P}^{(\mbox{\tiny{TOT}})}$ and the normal ${\hat {\bm n}}$ at the non-ideal interface
\begin{equation}
\tilde{\bm P}^{(\mbox{\tiny{TOT}})}={\bm P}^{(\mbox{\tiny{TOT}})}-{\bm P}^{(0)}=\left[P_{\rm b}+ \frac{3}{4} g_{AB} c_S^4 \rho_A {\Delta} \rho_B+ \frac{3}{4} g_{AB} c_S^4 \rho_B {\Delta} \rho_A  \right]{\bm \delta}-g_{AB} c_S^4 |{\bm \nabla} \rho_A| |{\bm \nabla} \rho_B| \left( {\bm \delta}-{\hat {\bm n}} {\hat {\bm n}}  \right)
\end{equation}
where we have used ${\bm \nabla} \rho_A \cdot {\bm \nabla} \rho_B=-|{\bm \nabla} \rho_A| |{\bm \nabla} \rho_B|$ and also the property of the normal ${\hat {\bm n}}$
$$
{\hat {\bm n}} =-\frac{{\bm \nabla} \rho_A}{|{\bm \nabla} \rho_A|}=\frac{{\bm \nabla} \rho_B}{|{\bm \nabla} \rho_B|}
$$
which is taken to point from the $A$-rich into the $B$-rich phase. We then introduce the tensor 
$$
{\bm T}={\bm \delta} -{\hat {\bm n}} {\hat {\bm n}} 
$$  
with the property to project along the tangential direction ${\hat {\bm t}}$, perpendicularly to ${\hat {\bm n}}$, i.e. ${\bm T} \cdot  {\hat {\bm n}} =0$. For our curved interface, we can then compute the pressure tensor in the directions normal (N) and tangential (T) to the interface and obtain
\be\label{eq:1}
\tilde{P}_{N}= P_{\rm b}+\frac{3}{4} g_{AB} c_S^4 \left( \rho_A {\Delta} \rho_B+ \rho_B {\Delta} \rho_A \right)  \hspace{.2in} \tilde{P}_{T}=P_{\rm b}+ \frac{3}{4} g_{AB} c_S^4 \left( \rho_A {\Delta} \rho_B+ \rho_B {\Delta} \rho_A \right)- g_{AB} c_S^4 |{\bm \nabla} \rho_A| |{\bm \nabla} \rho_B|
\ee
so that
\be\label{eq:2}
\tilde{P}_N-\tilde{P}_{T}= g_{AB} c_S^4 |{\bm \nabla} \rho_A| |{\bm \nabla} \rho_B|
\ee
and the pressure tensor can be rewritten as~\cite{blockhuis}
\be\label{maindecomposition}
\tilde{\bm P}^{(\mbox{\tiny{TOT}})}=\tilde{P}_{N} {\bm \delta}-{\bm T} (\tilde{P}_N-\tilde{P}_{T}).
\ee
When analyzing the equilibrium properties (${\bm \nabla} \cdot \tilde{{\bm P}}^{(\mbox{\tiny{TOT}})}=0$) with this decomposition, the mechanical balance at the interface is better understood. Indeed we get
\be\label{eq:general_spherical}
{\bm \nabla} \cdot \tilde{\bm P}^{(\mbox{\tiny{TOT}})}={\hat {\bm n}} \, ({\nabla}_n \tilde{P}_N) +\frac{{\hat {\bm n}}}{R} (\tilde{P}_N-\tilde{P}_T)  - {\hat {\bm t}} \, {\nabla}_t (\tilde{P}_N-\tilde{P}_T)
\ee
where we have introduced the gradients along the tangential and normal direction, i.e. ${\nabla}_n={\hat {\bm n}} \cdot {\bm \nabla}$, ${\nabla}_t={\hat {\bm t}} \cdot {\bm \nabla}$. Further, we have used that ${\bm \nabla} \cdot {\bm T} = -\frac{1}{R} {\hat {\bm n}}$ and also assumed that the gradient of the density fields is perpendicular to the interface (i.e. directed along ${\hat {\bm n}}$). When we deal with cases with homogeneous and isotropic surface tension, Eq.~\eqref{eq:general_spherical} leads to the well known Laplace equation. To see this point, let us specialize our equations to the case of a cylindrical droplet described in polar coordinates, with the center of coordinates coinciding with the center of the droplet ($r=0$). Because of the cylindrical symmetry, we can neglect the term  ${\hat {\bm t}}\, {\nabla}_t (\tilde{P}_N-\tilde{P}_T)$ to obtain
\be\label{BALANCEpart}
{\bm \nabla} \cdot \tilde{{\bm P}}^{(\mbox{\tiny{TOT}})}={\hat {\bm n}} \, ( {\nabla}_n \tilde{P}_N ) + \frac{{\hat {\bm n}}}{R} (\tilde{P}_N-\tilde{P}_T)   =0.
\ee
Integrating in the normal (i.e. radial) direction, we get
$$
P_{\rm b}^{(\mbox{\tiny{out}})}-P_{\rm b}^{(\mbox{\tiny{in}})}+\frac{1}{R} \int_0^{\infty} (\tilde{P}_N-\tilde{P}_T) \, d r=0 
$$
that is the usual Laplace law written for the bulk pressure $P_{\rm b}$ evaluated in the outer (out, $r=\infty$) and inner (in, $r=0$) regions. The surface tension $\sigma$ follows 
$$
\sigma=\int_0^{\infty} (\tilde{P}_N-\tilde{P}_T) \, d r.
$$
In order to construct a force driven by gradients in the surface tension, we introduce ``on the top'' of the total pressure tensor $\tilde{\bm P}^{(\mbox{\tiny{TOT}})}$ a temperature-dependent term. In particular, following~\cite{SbragagliaGuptaScagliarini}, and relying on the MRT schemes, we have added the new term to the stress modes of the multicomponent model, in such a way that the resulting hydrodynamic Eqs~\eqref{NScd} with the coupling~\eqref{eq:additionalcoupling} are recovered in the limit of a thin interface. The choice for the temperature-dependent term is quite phenomenological~\cite{Zhangetal12,Zhangetal13,Zhangetal14,Li15}: it is chosen to be proportional to the squared gradient of the ``order parameter'', the latter taken as the density difference $\rho_d=\rho_A-\rho_B$, and is also proportional to a temperature-dependent prefactor
$$
\tilde{{\bm P}}^{(\mbox{\tiny{TOT}})} \rightarrow \tilde{{\bm P}}^{(\mbox{\tiny{TOT}})}-c_S^4 A(T) |{\bm \nabla} \rho_d|^2{\bm T}.
$$
From~\eqref{eq:1}-\eqref{eq:2}, it follows that the difference between the normal and tangential component of the pressure tensor changes as $(\tilde{P}_N-\tilde{P}_T) \rightarrow (\tilde{P}_N-\tilde{P}_T) + c_S^4 A(T) |{\bm \nabla} \rho_d|^2$. Hence, the balance equation at a curved interface is modified as
\be\label{BALANCE2}
{\bm \nabla} \cdot \tilde{{\bm P}}^{(\mbox{\tiny{TOT}})}={\hat {\bm n}} \, \left( {\nabla}_n \tilde{P}_N \right) +\frac{{\hat {\bm n}}}{R} (\tilde{P}_N-\tilde{P}_T+c_S^4 A(T)|{\bm \nabla} \rho_d|^2)  - {\hat {\bm t}} \, {\nabla}_t (\tilde{P}_N-\tilde{P}_T+c_S^4 A(T)|{\bm \nabla} \rho_d|^2) 
\ee
that is producing a temperature dependent surface tension
$$
\sigma (T)=\int_0^{\infty} (P_N-P_T) \, d r+ \int_0^{\infty} c_S^4 A(T)|{\bm \nabla} \rho_d|^2 \, d r
$$
while the term ${\hat {\bm t}} \, {\nabla}_t (\tilde{P}_N-\tilde{P}_T+c_S^4 A(T)|{\bm \nabla} \rho_d|^2)$  is tangential at the interface and is responsible for the Marangoni effects triggered by the tangential variation of the surface tension $\sigma(T)$~\cite{Zhangetal12,Zhangetal13,Zhangetal14,Li15}. The simplest choice for $A(T)$ reproducing the motion of a curved interface from cold to hot regions is~\cite{Zhangetal12,Zhangetal13,Zhangetal14,Li15} 
\be\label{EFFEKAPPA}
A(T)=G_0-\GT T
\ee 
where $G_0$ is a constant and $\GT>0$. With this choice, the surface tension is larger in regions where the temperature is lower. \\

\subsection{Modelling of Advection-Diffusion Equation for the Temperature Field}\label{sec:LBTHERMAL}

For the temperature dynamics, following~\cite{Ripesi14}, we add another set of populations $g_{s} ({\bm{x}}, t)$. At difference with the multicomponent case, instead of using the MRT approach, the collisional part of the LB dynamics is now taken to be a BGK relaxation dynamics for simplicity
\begin{equation}\label{EQ:LBappg}
g_{s} ({\bm{x}} + {\bm{c}}_{s} , t + 1)-g_{s} ({\bm{x}}, t) = -\frac{1}{\tau_g}(g_{s}-g^{(eq)}_{s}) 
\end{equation}
where $\tau_g$ is a characteristic time governing the relaxation dynamics towards the local equilibrium distribution
\be\label{feq}
g_{s}^{(eq)}=w_{s} T \left[1+\frac{{\bm{u}} \cdot {\bm{c}}_{s}}{c_S^2}+\frac{{\bm{u}}{\bm{u}}:({\bm{c}}_{s}{\bm{c}}_{s}-{\bm \delta})}{2 c_S^4} \right].
\ee
As in the case of the multicomponent part of the model, the hydrodynamic evolution for the temperature results from the long-wavelength limit~\cite{Gladrow00,Succi01} of the equations for the coarse-grained temperature fields, the latter computed as the ``density'' of the $g_s$ population, i.e. $T=\sum_s g_s$. One can show that Eqs~\eqref{THERMAL} are recovered with $\kappa^{(T)}_{\rm c,d}=c_S^2 (\tau_g-1/2)$.  Let us also note that, in order to get the right hydrodynamic limit of the temperature evolution, an extra body-force term is in principle needed in order to avoid spurious terms in the continuum limit~\cite{Latt07}. The importance of this term may depend on the applications. In all our simulations, we have checked that it is negligible. Notice that for more complex cases where thermal effects may enter the equation of state, one can also devise different strategies~\cite{Scagliarini10,Biferale13}. 

\section{T-Junction set-up}\label{sec:TJsetup}

In this section we report the main details for the T-junction set-up used for the numerical simulations. The T-junction is embedded in a rectangular parallelepiped with size $L_{x}  \times L_y \times H$, and both the main and side channels have a square cross-section resolved with $H \times H = 32 \times 32$ lattice cells. The resolution used for $H$ is chosen to be compatible with those used for the successful benchmark tests described in Appendix~\ref{sec:benchmark}. The main channel and the side channel lengths are resolved with a $L_x=640$ lattice cells and $L_y=160$ lattice cells, respectively. All the simulation details are summarized in table~\ref{table:para}. In order to highlight the thermocapillary effects in the transition from squeezing to dripping, at changing a variety of flow-rates in both the continuous and dispersed phases, we decided to keep the viscous ratio $\lambda=\etad/\etac$ and thermal diffusivity ratio $\chi=\kappa^{(T)}_{\rm c}/\kappa^{(T)}_{\rm d}$ fixed to unity. Moreover, the Reynolds number, $\Ren =\rho \vc H/\etac$, is small, hence it is expected a minor influence on the thread deformation and break-up~\cite{RenardyCristini01}.  We then choose the following dimensionless numbers~\cite{Demenech07,LiuZhang09,LiuZhang11} to fully characterize the system. First, the Capillary number $\Ca$ that is calculated in the continuous phase
\be\label{eq:Ca}
\Ca = \frac{\etac \vc}{\sigma}
\ee 
and the flow-rate ratio
\be\label{eq:Q}
\phi=\frac{\vd}{\vc} = \frac{\Qd}{\Qc}
\ee
where $\Qd=\vd H^2$ and $\Qc =\vc H^2$ are the flow-rates at the two inlets of the side and main channel. We emphasize that the flow-rates are input parameters for the T-Junction set-up. In addition to the two control parameters $\Ca$ and $\phi$, to quantify the effects of thermocapillarity, we use the Marangoni number
\be\label{eq:Ma}
\Ma=\frac{H \Delta \sigma_{H} }{\kappac^{(T)} \etac}
\ee
where $\Delta \sigma_{H}$ represents the variation in the surface tension over a distance $H$ (see also Appendix~\ref{sec:Appendix}). In the outlet, we impose a fixed pressure and also use Neumann-type boundary conditions for the velocity fields. A Dirichlet-type boundary condition is set at the inlets of the main and side channel, by imposing the pressure gradient computed from the exact solution of a Stokes flow in a square duct~\cite{vandersman08}. As for the temperature boundary conditions, we set a constant temperature at the inlet ($T=T_{\mbox{\tiny{inlet}}}$) and outlet ($T=T_{\mbox{\tiny{outlet}}}$) of the main channel. In contact with the walls of the main and side channels, we set a Dirichlet-type boundary condition for the temperature field that is a space-dependent one: it coincides with the linear profile going from $T_{\mbox{\tiny{inlet}}}$ to $T_{\mbox{\tiny{outlet}}}$. In the inlet of the side channel we set the temperature to be consistent with such linear profile. By suitably changing the parameter $\GT>0$ in~\eqref{EFFEKAPPA}, while setting $T_{\mbox{\tiny{inlet}}}-T_{\mbox{\tiny{outlet}}}=\pm 1.4$ lbu, we are therefore able to explore situations where the temperature gradient promotes migration in the positive/negative main channel direction, dependently on the sign of the temperature gradient ${\nabla} T=(T_{\mbox{\tiny{inlet}}}-T_{\mbox{\tiny{outlet}}})/L_x$ (see also Fig.~\ref{fig:Ca_001_rho1}). Notice that we use the same temperature gradient that we tested in the benchmark tests in Appendix~\ref{sec:benchmark}, hence we already know the characteristic thermocapillary velocities developed in such temperature gradients (see Fig.~\ref{fig:2}). Moreover, in presence of a non zero temperature gradient, the surface tension $\sigma=\sigma(T)$ used to compute the Capillary number~\eqref{eq:Ca} is the one corresponding to the temperature at the T-junction \footnote{We take the temperature in the center of the junction.}, i.e., where the break-up process takes place. In addition, wetting properties need to be introduced at the boundaries. To this aim, we impose perfect wetting for the continuous phase.

\begin{table*}[t!]

\begin{minipage}[c]{0.5\textwidth}%
\begin{center}
\includegraphics[width = 1.0\linewidth]{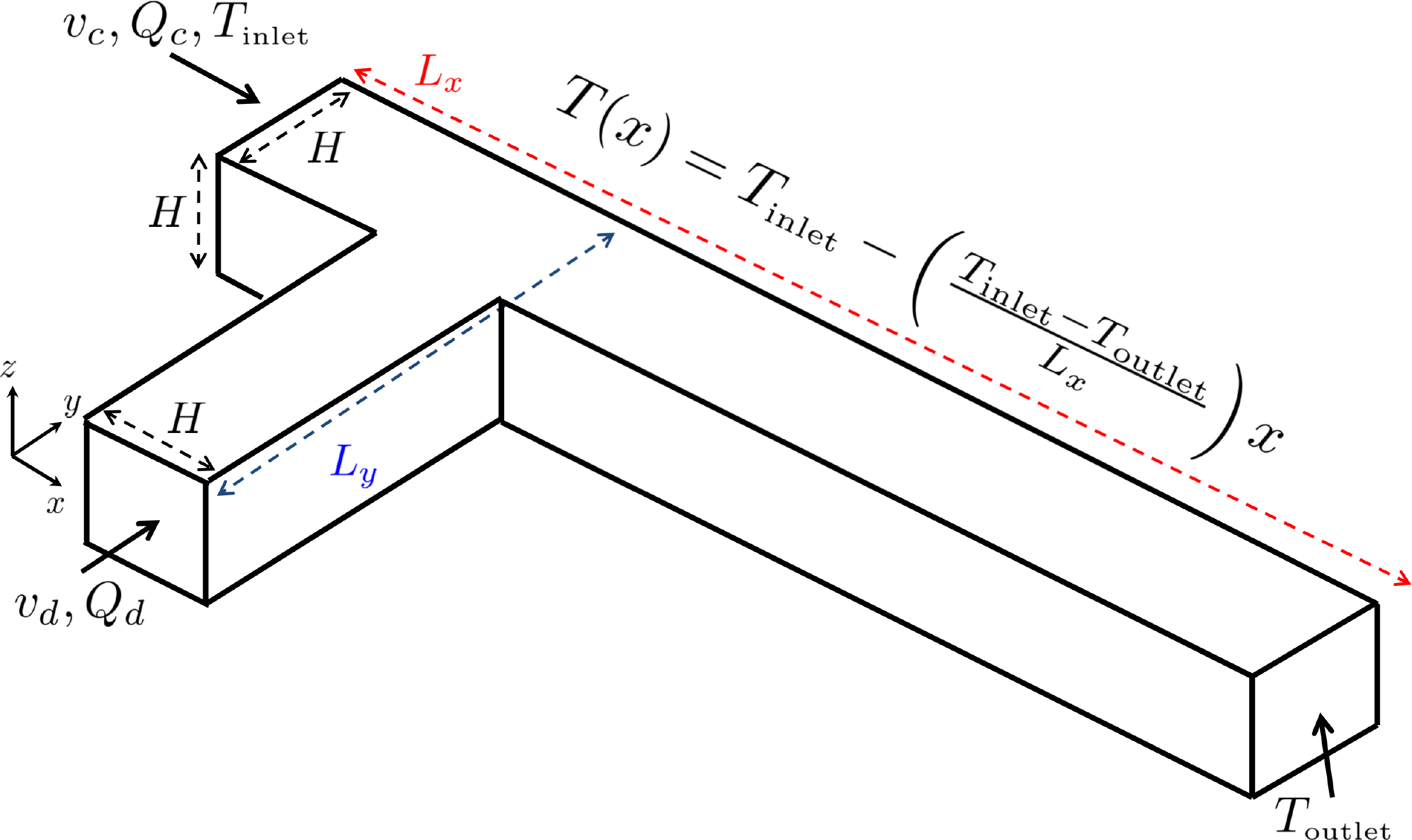}
\end{center}
\end{minipage}

\vspace{.4in}
\parbox{1.0\textwidth}{
   \begin{tabular}{@{\extracolsep{\fill}} |c|c|c|c|c|c|c|c|c|c|c|c|}
    \hline
    $\Ca$ & $\phi$ & $\Ma$ & $\nabla T$ & $L_{x} \times L_y \times H$ & $H$ & $\etad$ & $\etac$  & $\lambda$ & $\kappac^{(T)}$ & $\kappad^{(T)}$ & $\chi$ \\
   & & & & cells & cells & lbu & lbu & & lbu & lbu & \\
   \hline \hline
    $0.001-0.1$ & $ 0.2-1.0$ & $0.0$ & $=0$ & $640 \times 160 \times 32$ & $32$ & $0.28$ & $0.28$ & $1.0$ & $0.166$ & $0.166$ & $1.0$\\
   \hline \hline
    $0.001-0.1$ & $ 0.2-1.0$ & $0.378-1.26$ & $>0$ & $640 \times 160 \times 32$ & $32$ & $0.28$ & $0.28$ & $1.0$ & $0.166$ & $0.166$ & $1.0$\\
   \hline 
   \hline 
    $0.001-0.1$ & $ 0.2-1.0$ & $0.378-1.26$ & $<0$ & $640 \times 160 \times 32$ & $32$ & $0.28$ & $0.28$ & $1.0$ & $0.166$ & $0.166$ & $1.0$\\
\hline
   \end{tabular}
}
\caption{Parameters for the lattice Boltzmann (LB) simulations with the T-junction geometry. A rectangular parallelepiped with size $L_{x}  \times L_y \times H$ embeds the T-junction geometry. Both the main and side channels have square cross section ($H \times H$) with $H$ resolved with 32 lattice cells. The flow-rates at the two inlets are indicated with $\Qd=\vd H^2$ and $\Qc =\vc H^2$, where $v_{\rm d,c}$ represent the average velocity in the two phases. The flow-rate ratio is indicated with $\phi=\Qd/\Qc$. The dynamic viscosity of the dispersed phase is $\etad$, while $\etac$ is the dynamic viscosity of the continuous phase. The Capillary number $\Ca$ is computed in the continuous phase, $\Ca=\etac \vc/\sigma$, with $\sigma$ the surface tension. Notice that in presence of Marangoni effects, we use the surface tension $\sigma=\sigma(T)$ corresponding to the temperature at the junction, where break-up takes place. The advection-diffusion equation for the temperature field~\eqref{THERMAL} is integrated with an LB solver (see section~\ref{sec:LBTHERMAL}) with diffusivities $\kappa^{(T)}_{\rm d,c}$ in the dispersed and continuous phases, respectively. Notice that in order to perform different numerical simulations at changing $\Ca$ and $\phi$, we decided to use a unitary viscous ratio, $\lambda=\etad/\etac=1$, and unitary thermal diffusivity ratio, $\chi=\kappac^{(T)}/\kappad^{(T)}=1$. As for the temperature boundary conditions, a linear profile from the inlet ($T_{\mbox{\tiny{inlet}}}$) and outlet ($T_{\mbox{\tiny{outlet}}}$) temperature is imposed on the channel walls (see text for details). \label{table:para}}
\end{table*}


\section{Results for T-Junction Geometry}\label{sec:TJresults}

\begin{figure*}[h!]
\begin{minipage}{0.32\textwidth}
\subfigure[{\scriptsize $\Ma = 0.0$}]{\includegraphics[trim=10 390 0 0 , clip, width = 1.00\linewidth]{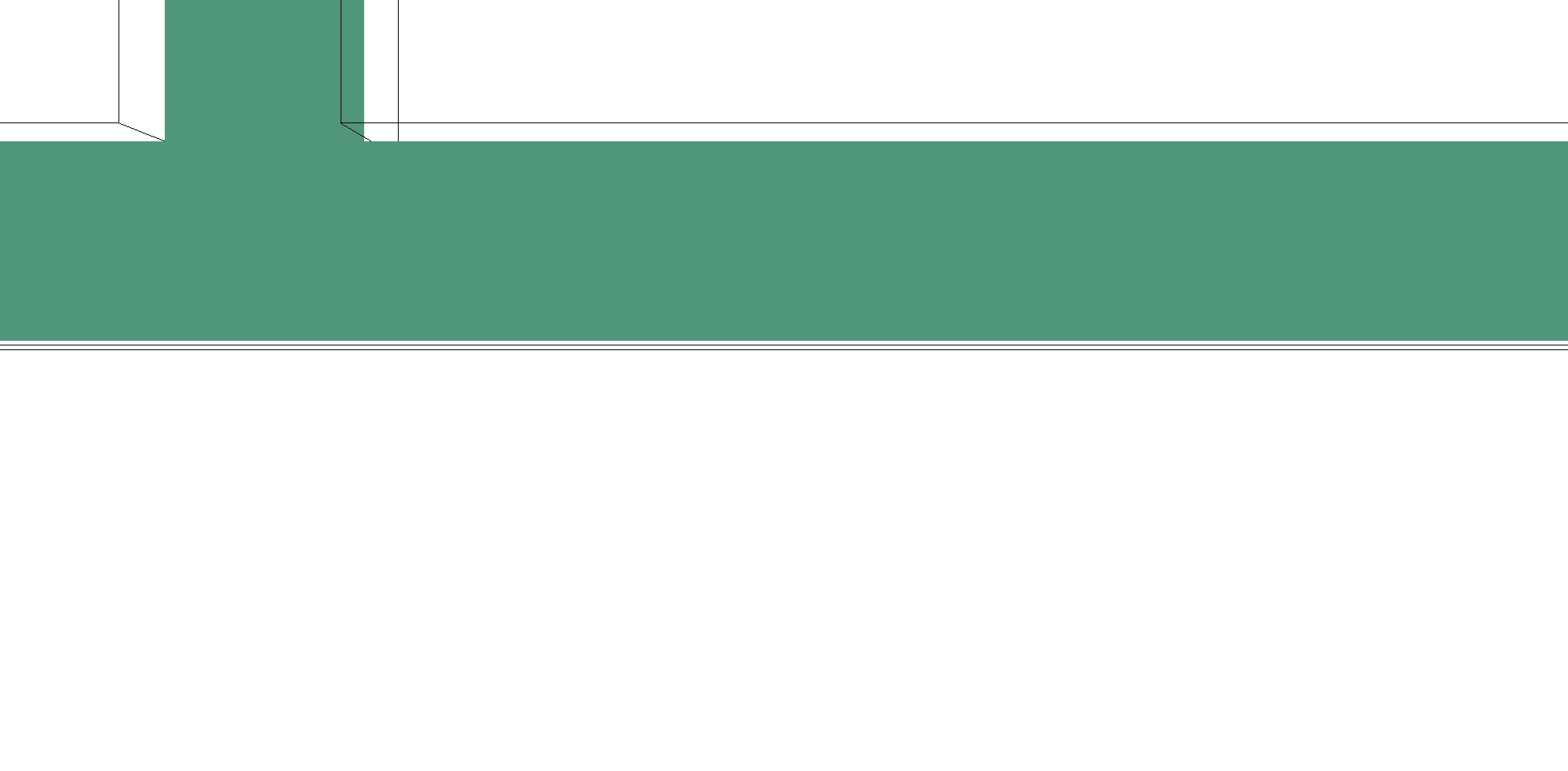}}\\
\subfigure[\,{\scriptsize $t = t_0 + 1.875 \, \tshear; \Ma = 0.0$}]{\includegraphics[trim=10 390 0 0 , clip, width = 1.00\linewidth]{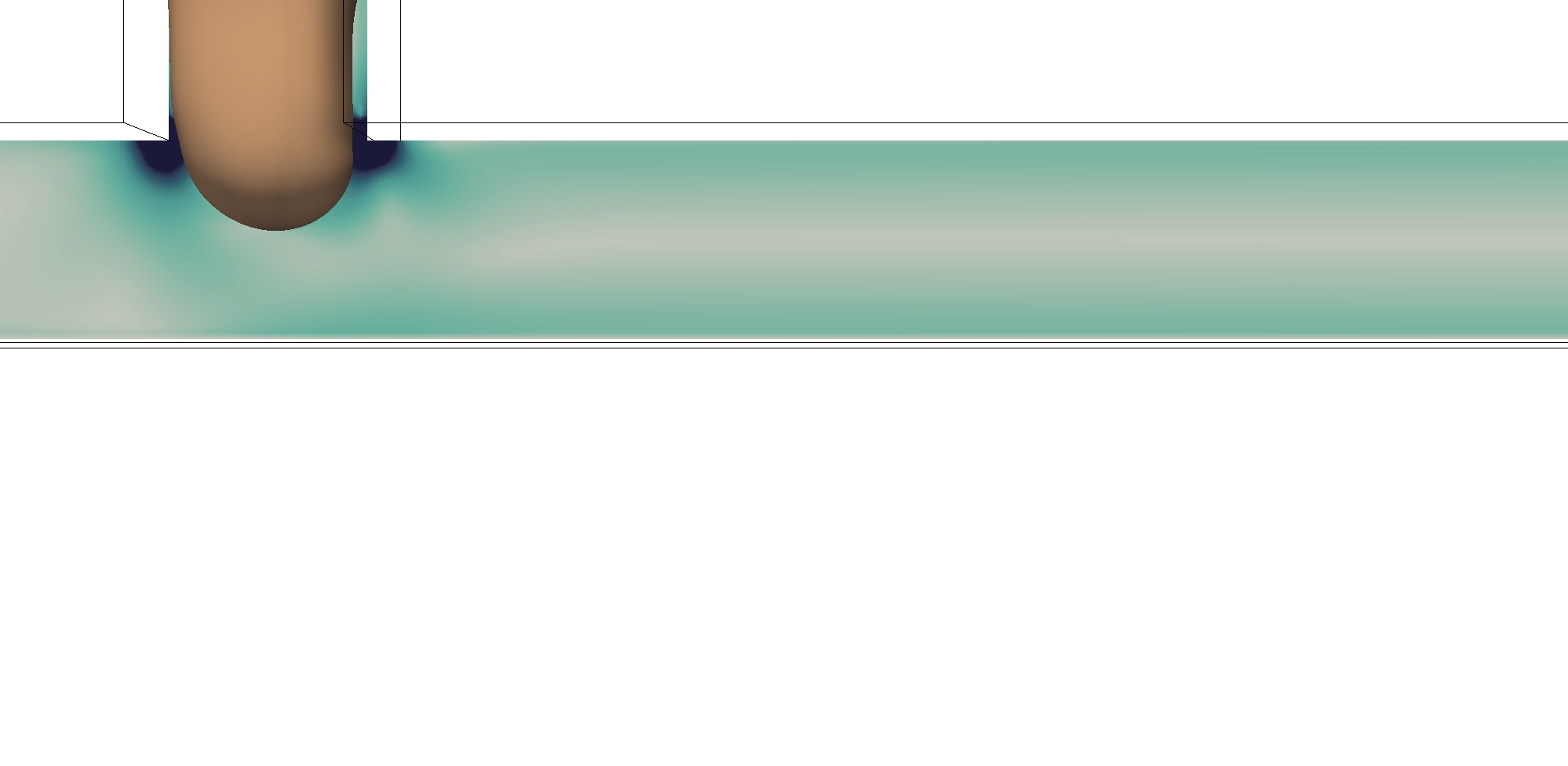}}\\
\subfigure[\,{\scriptsize $t = t_0 + 3.750 \, \tshear; \Ma = 0.0$}]{\includegraphics[trim=10 390 0 0 , clip, width = 1.00\linewidth]{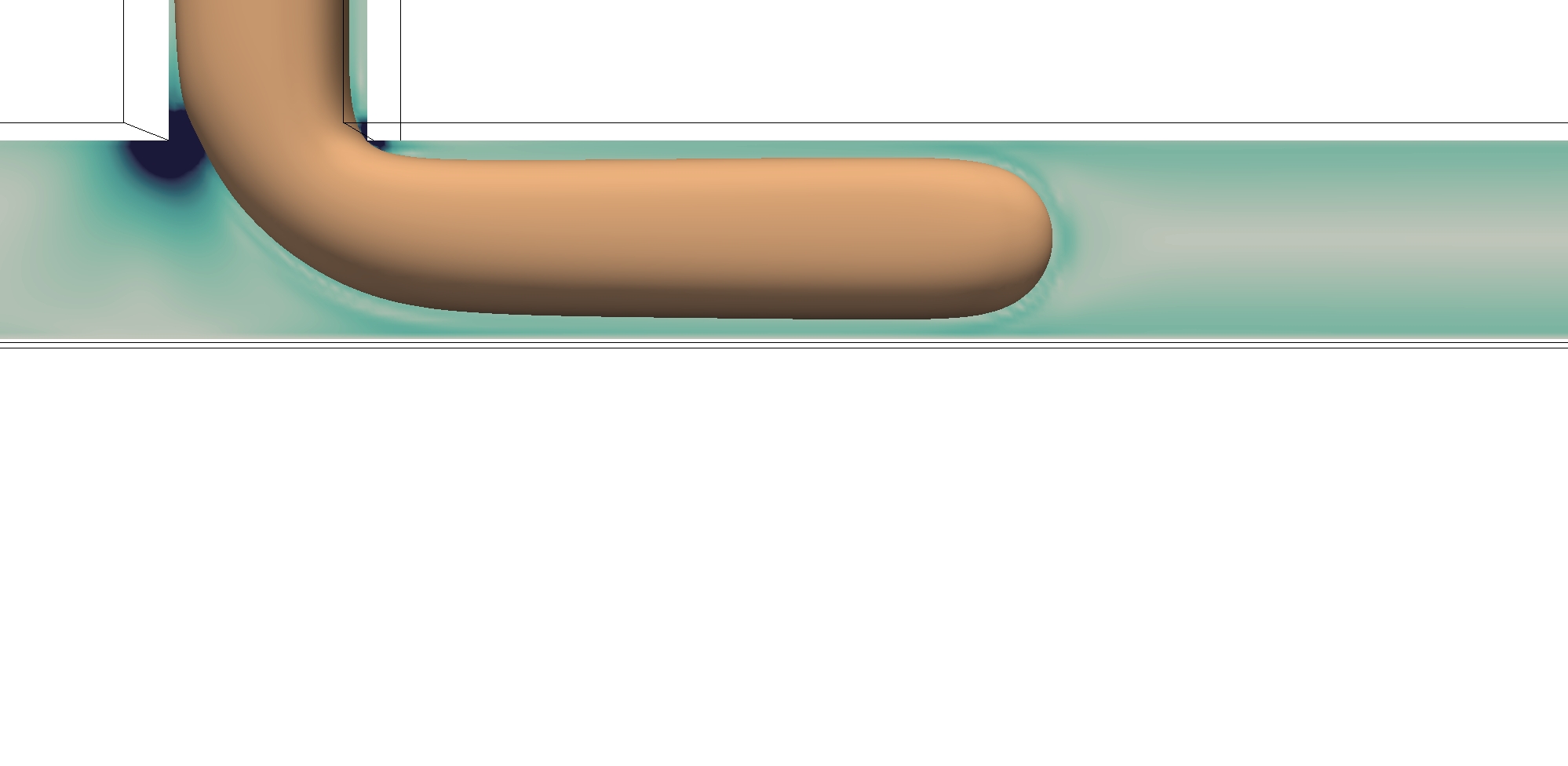}}\\
\subfigure[\,{\scriptsize $t = t_0 + 4.875 \, \tshear; \Ma = 0.0$}]{\includegraphics[trim=10 390 0 0 , clip, width = 1.00\linewidth]{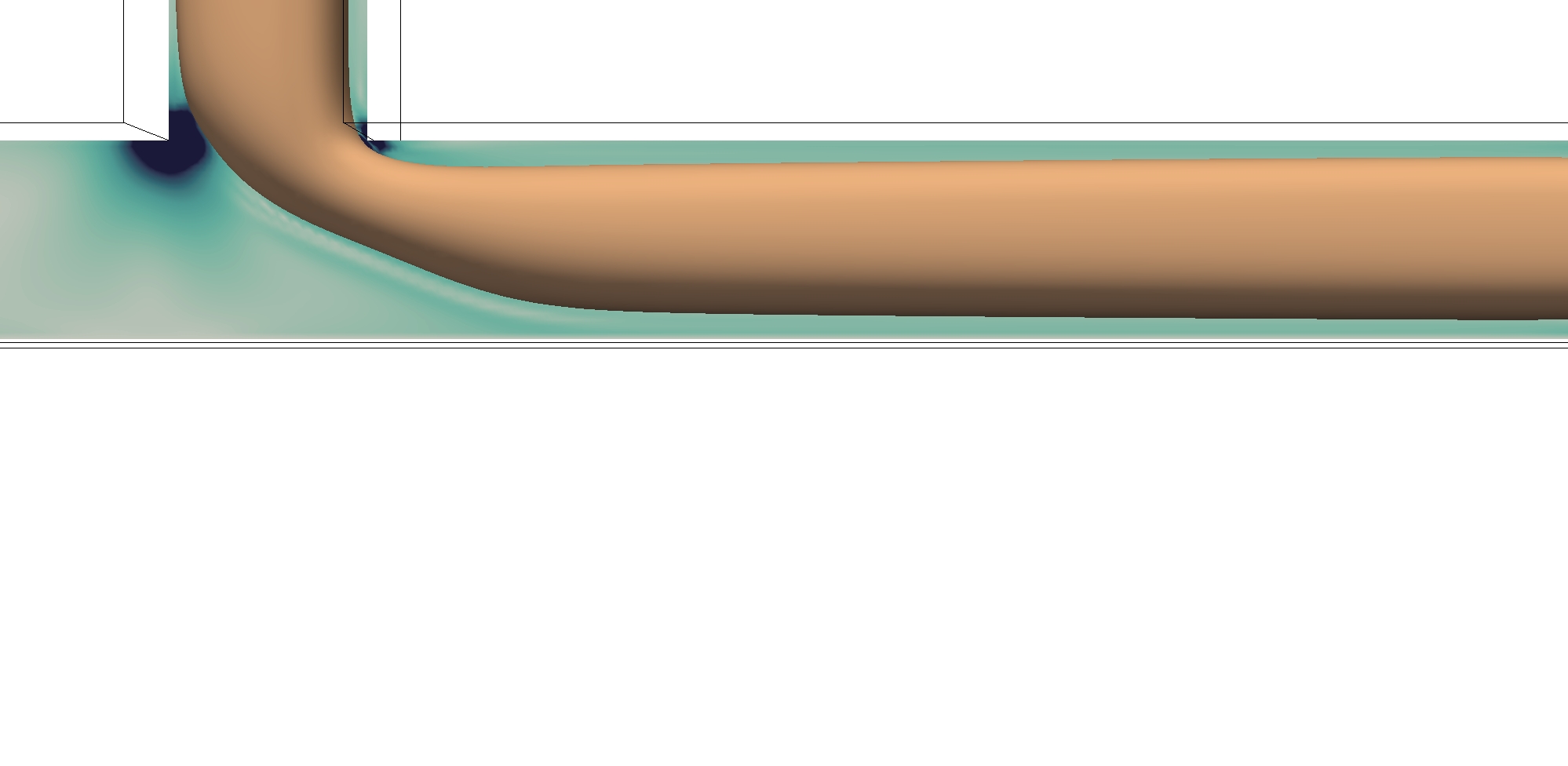}}\\
\subfigure[\,{\scriptsize $t = t_0 + 5.250 \, \tshear; \Ma = 0.0$}]{\includegraphics[trim=10 390 0 0 , clip, width = 1.00\linewidth]{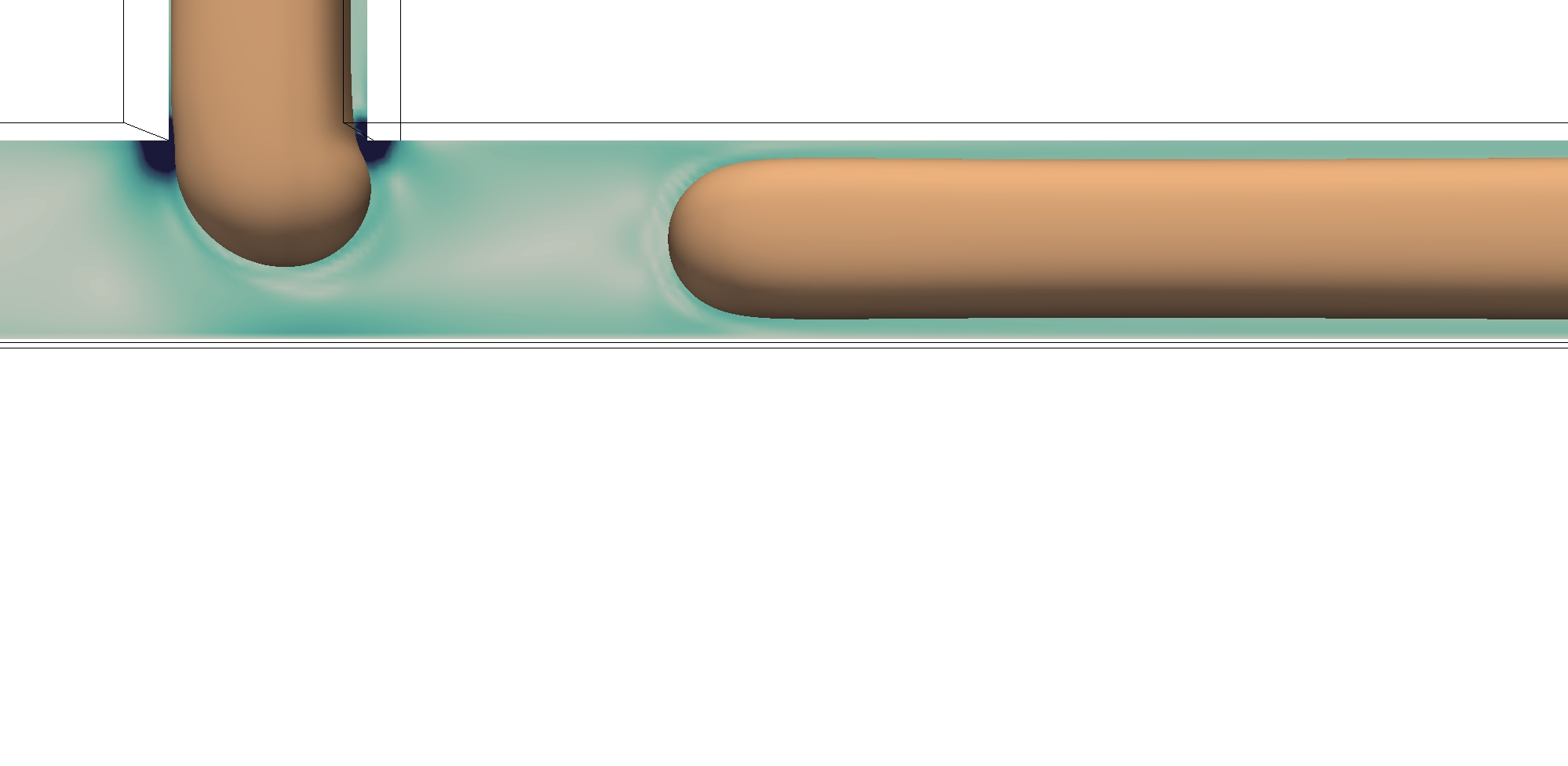}}
\end{minipage}
\begin{minipage}{0.32\textwidth}
\subfigure[\,{\scriptsize $\Ma = 1.26; \nabla T > 0$}]{\includegraphics[trim=10 390 0 0 , clip, width = 1.00\linewidth]{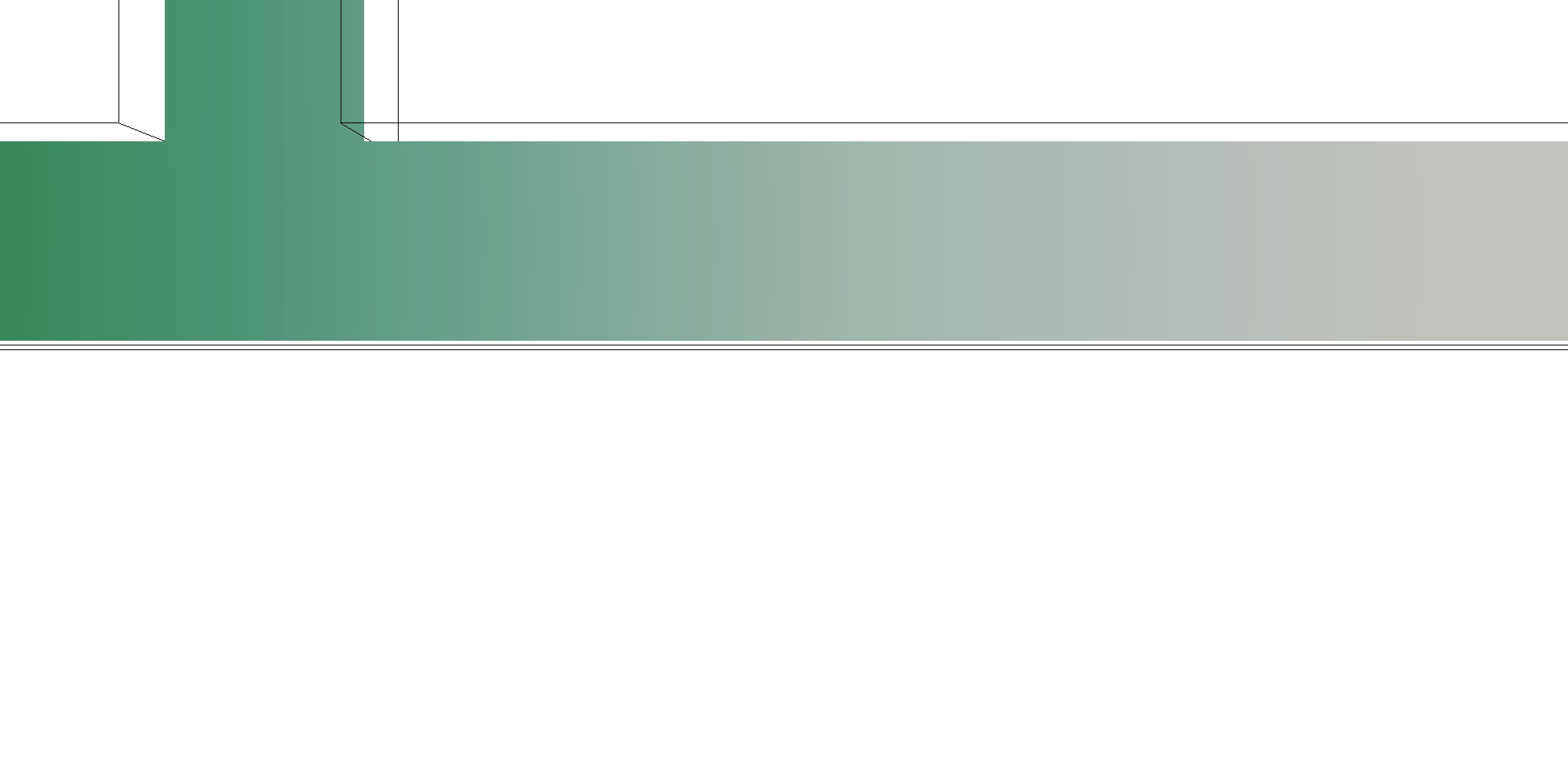}}\\
\subfigure[\,{\scriptsize $t = t_0 + 1.875 \, \tshear; \Ma = 1.26; \nabla T > 0$}]{\includegraphics[trim=10 390 0 0 , clip, width = 1.00\linewidth]{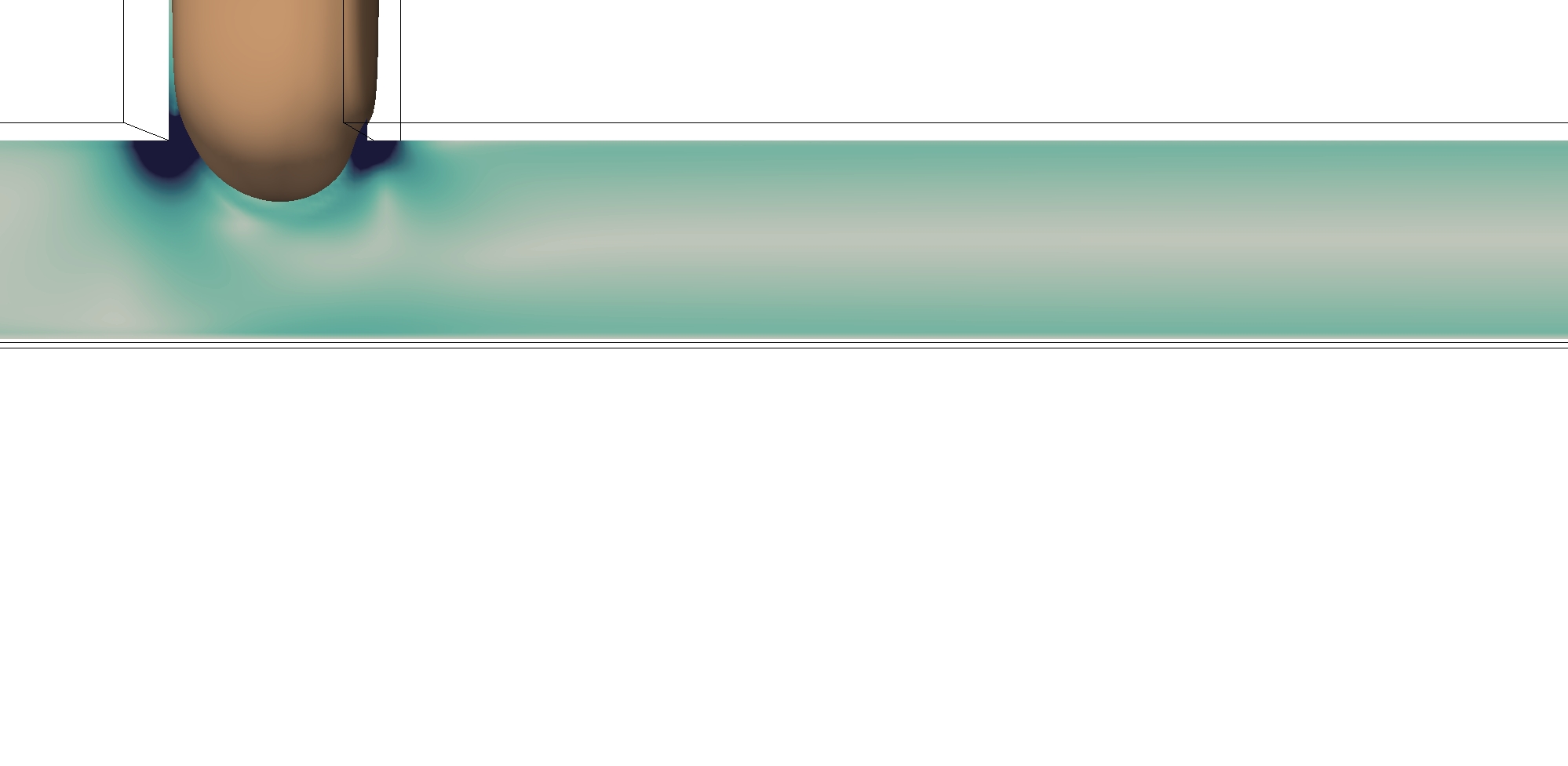}}\\
\subfigure[\,{\scriptsize $t = t_0 + 3.750 \, \tshear; \Ma = 1.26; \nabla T > 0$}]{\includegraphics[trim=10 390 0 0 , clip, width = 1.00\linewidth]{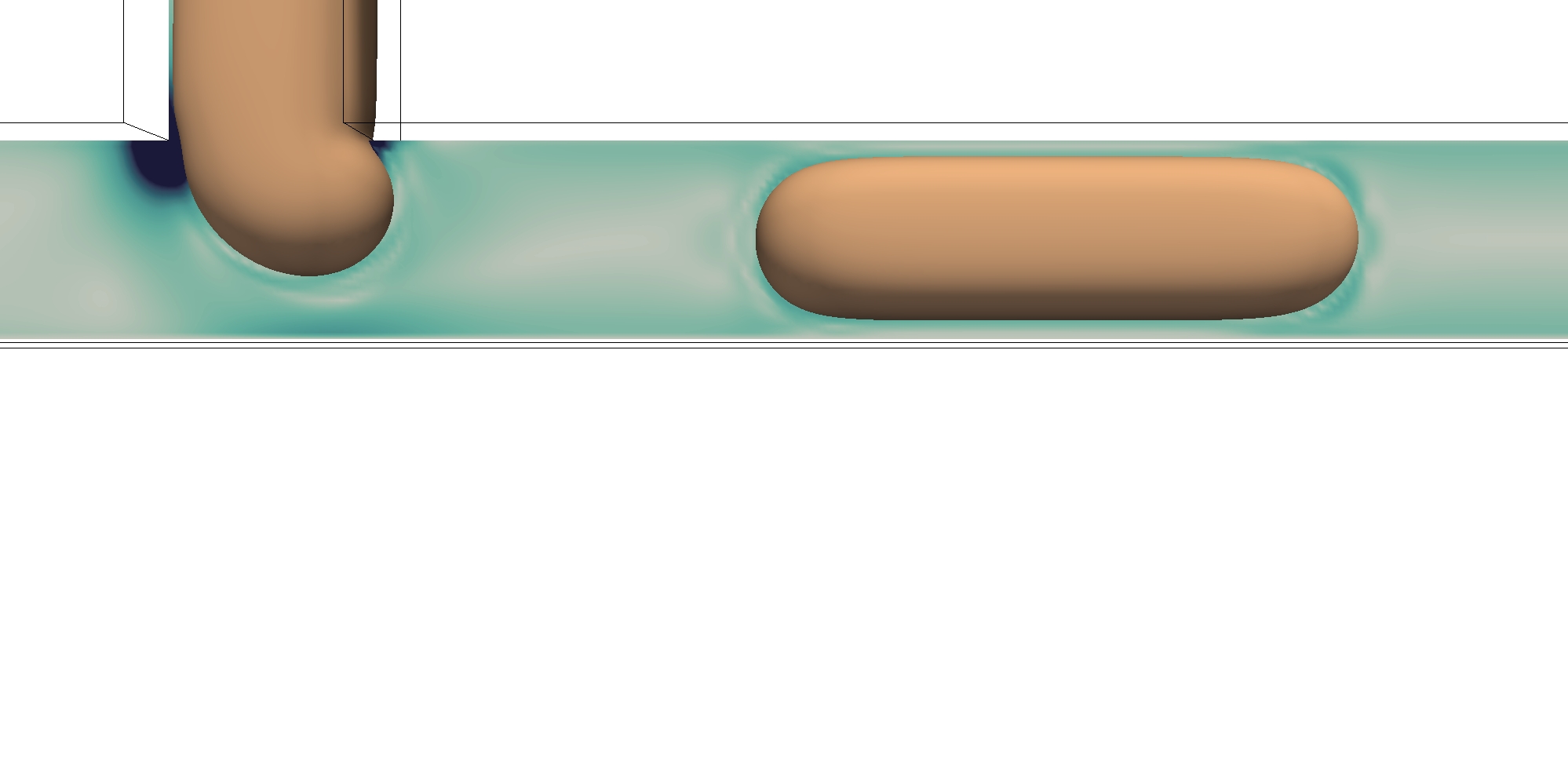}}\\
\subfigure[\,{\scriptsize $t = t_0 + 4.875 \, \tshear; \Ma = 1.26; \nabla T > 0$}]{\includegraphics[trim=10 390 0 0 , clip, width = 1.00\linewidth]{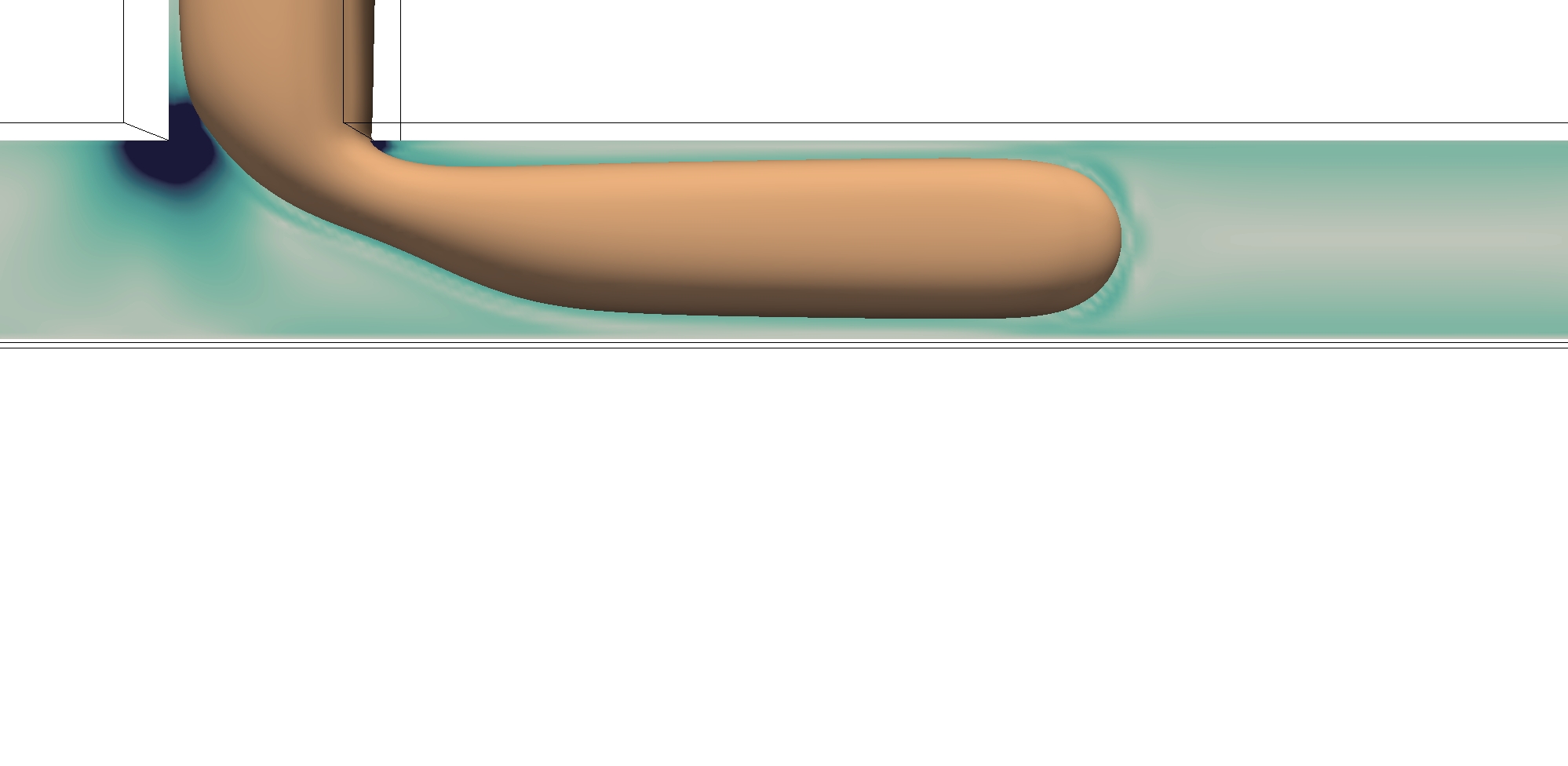}}\\
\subfigure[\,{\scriptsize $t = t_0 + 5.250 \, \tshear; \Ma = 1.26; \nabla T > 0$}]{\includegraphics[trim=10 390 0 0 , clip, width = 1.00\linewidth]{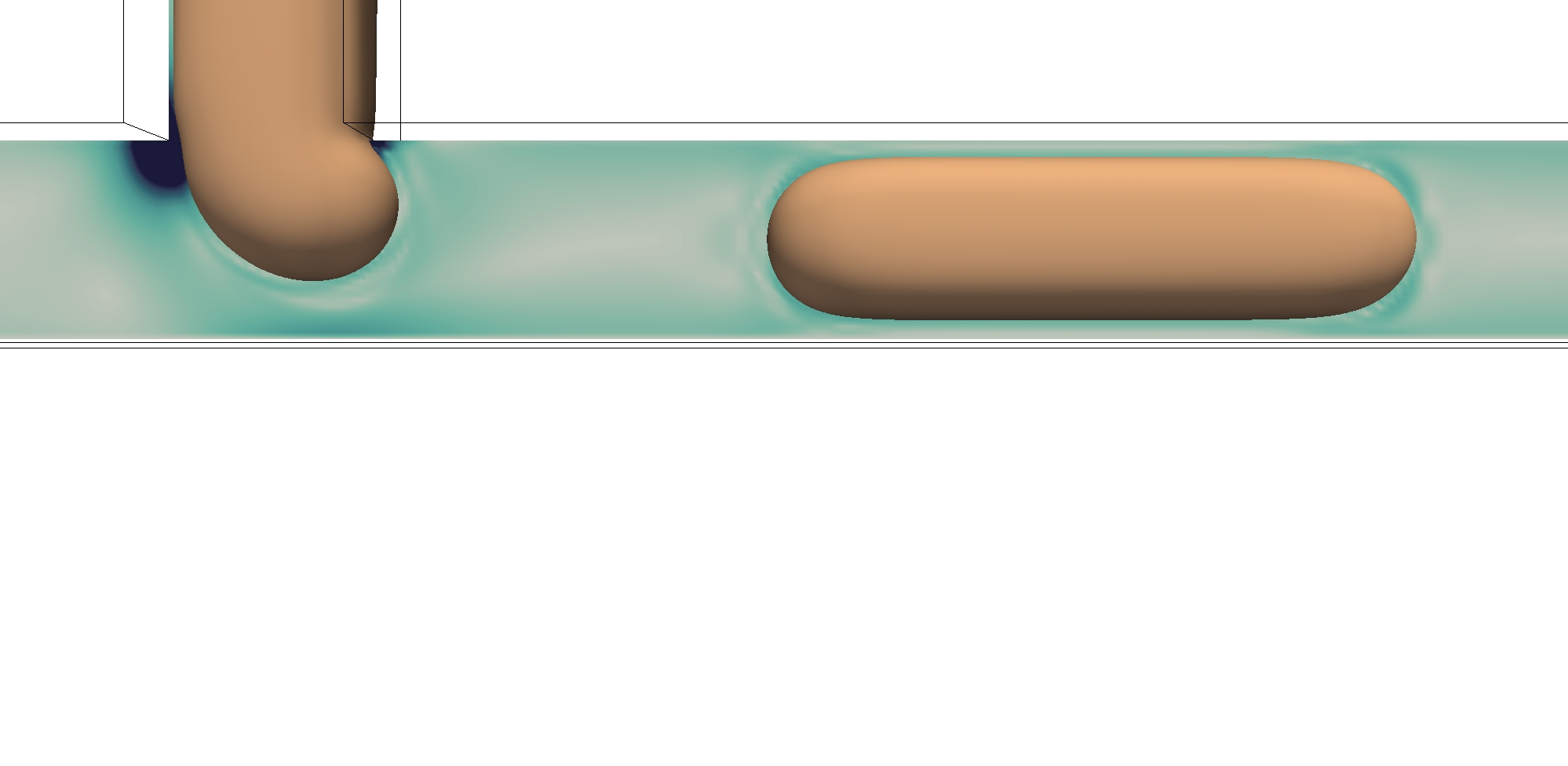}}
\end{minipage}
\begin{minipage}{0.32\textwidth}
\subfigure[\,{\scriptsize $\Ma = 1.26; \nabla T < 0$}]{\includegraphics[trim=10 390 0 0 , clip, width = 1.00\linewidth]{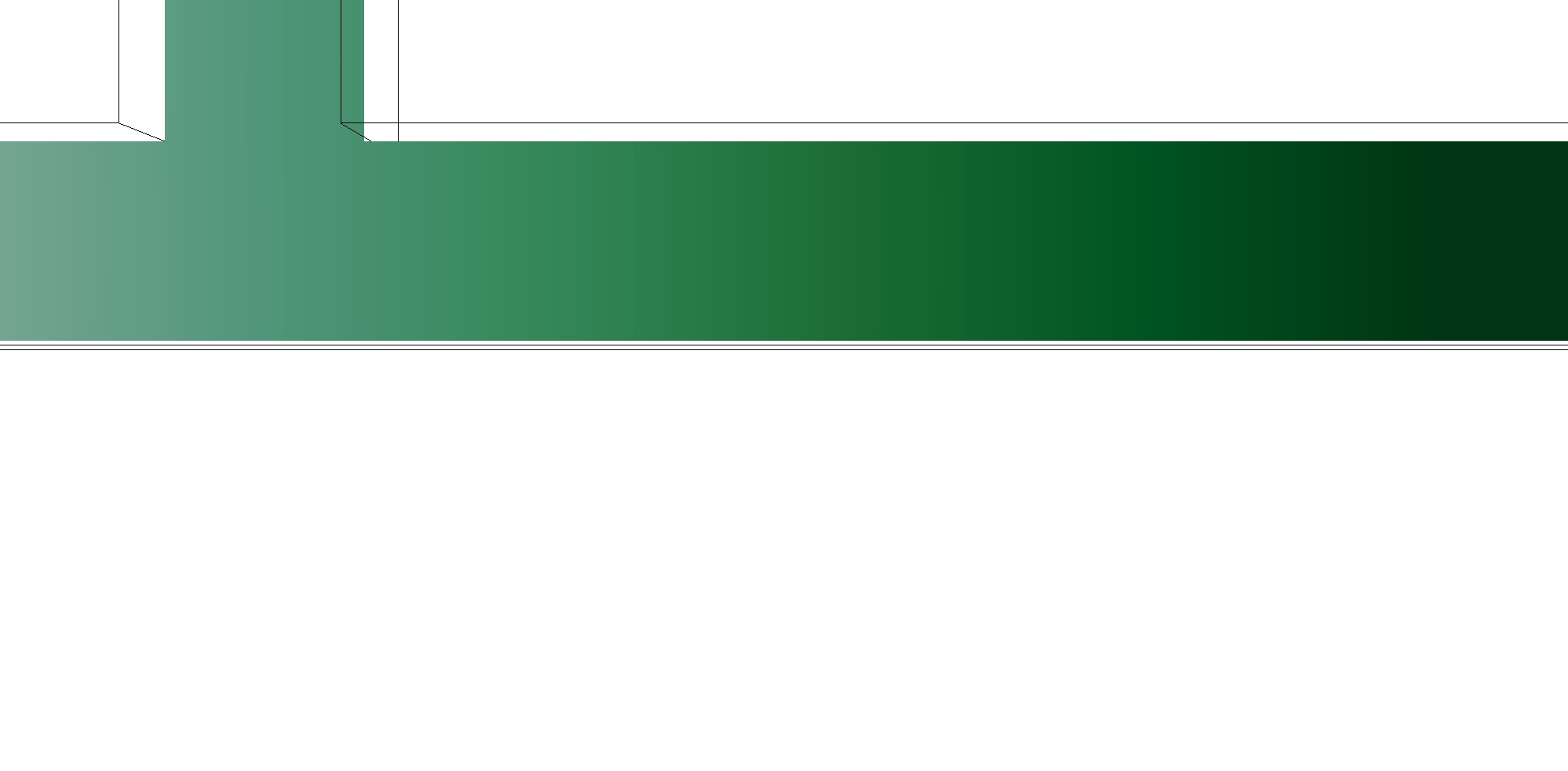}}\\
\subfigure[\,{\scriptsize $t = t_0 + 1.875 \, \tshear; \Ma = 1.26; \nabla T < 0$}]{\includegraphics[trim=10 390 0 0 , clip, width = 1.00\linewidth]{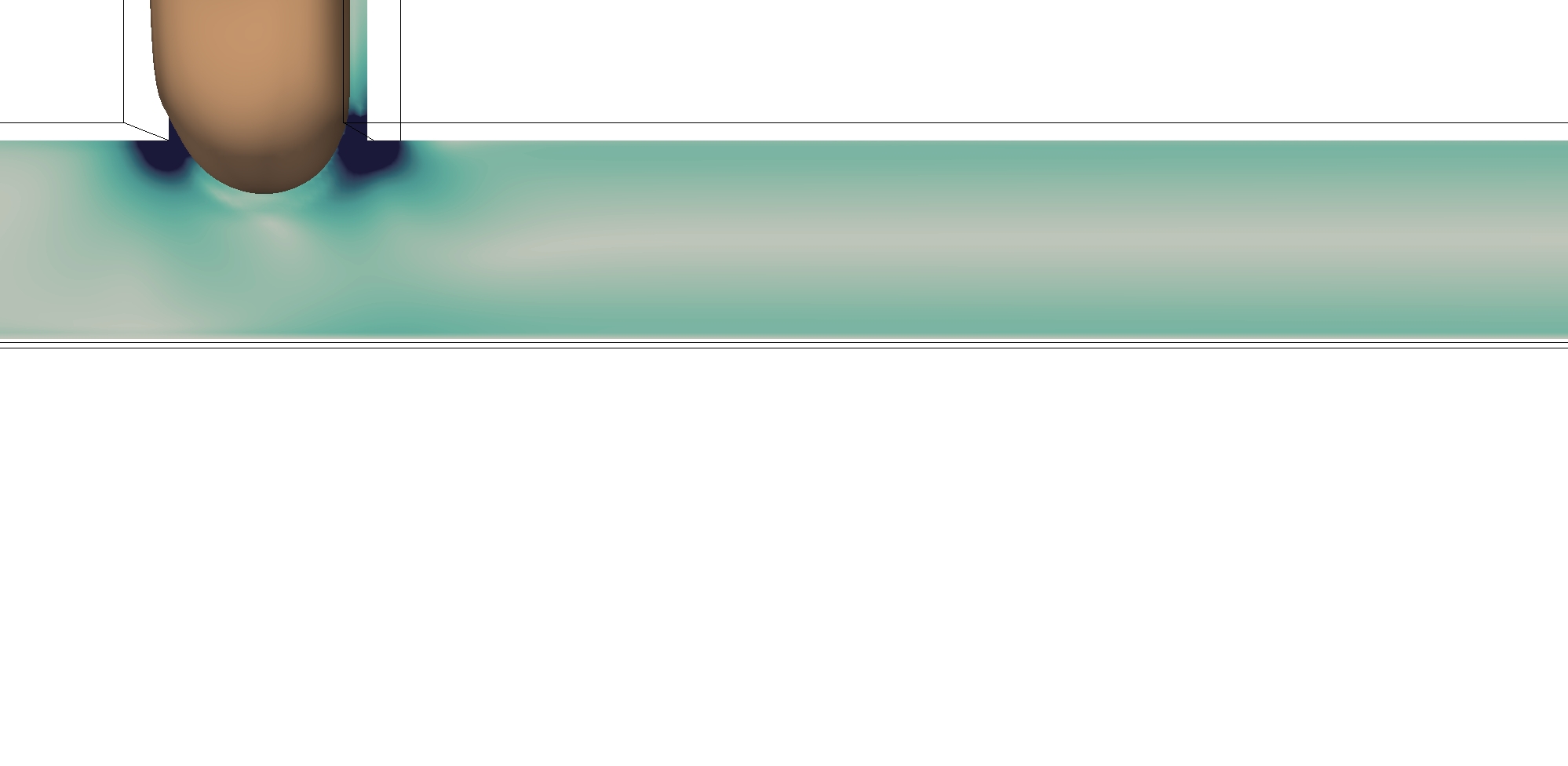}}\\
\subfigure[\,{\scriptsize $t = t_0 + 3.750 \, \tshear; \Ma = 1.26; \nabla T < 0$}]{\includegraphics[trim=10 390 0 0 , clip, width = 1.00\linewidth]{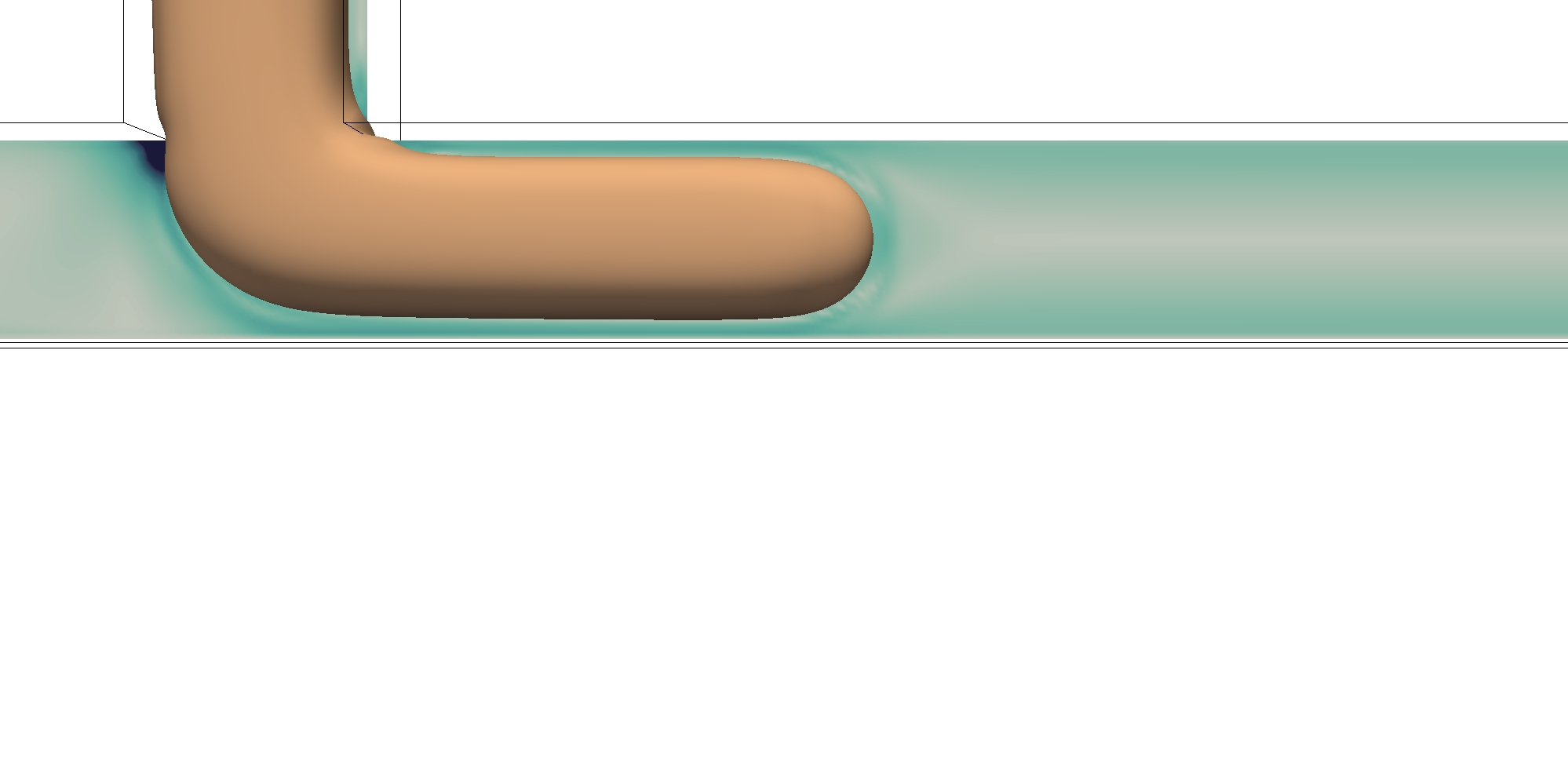}}\\
\subfigure[\,{\scriptsize $t = t_0 + 4.875 \, \tshear; \Ma = 1.26; \nabla T < 0$}]{\includegraphics[trim=10 390 0 0 , clip, width = 1.00\linewidth]{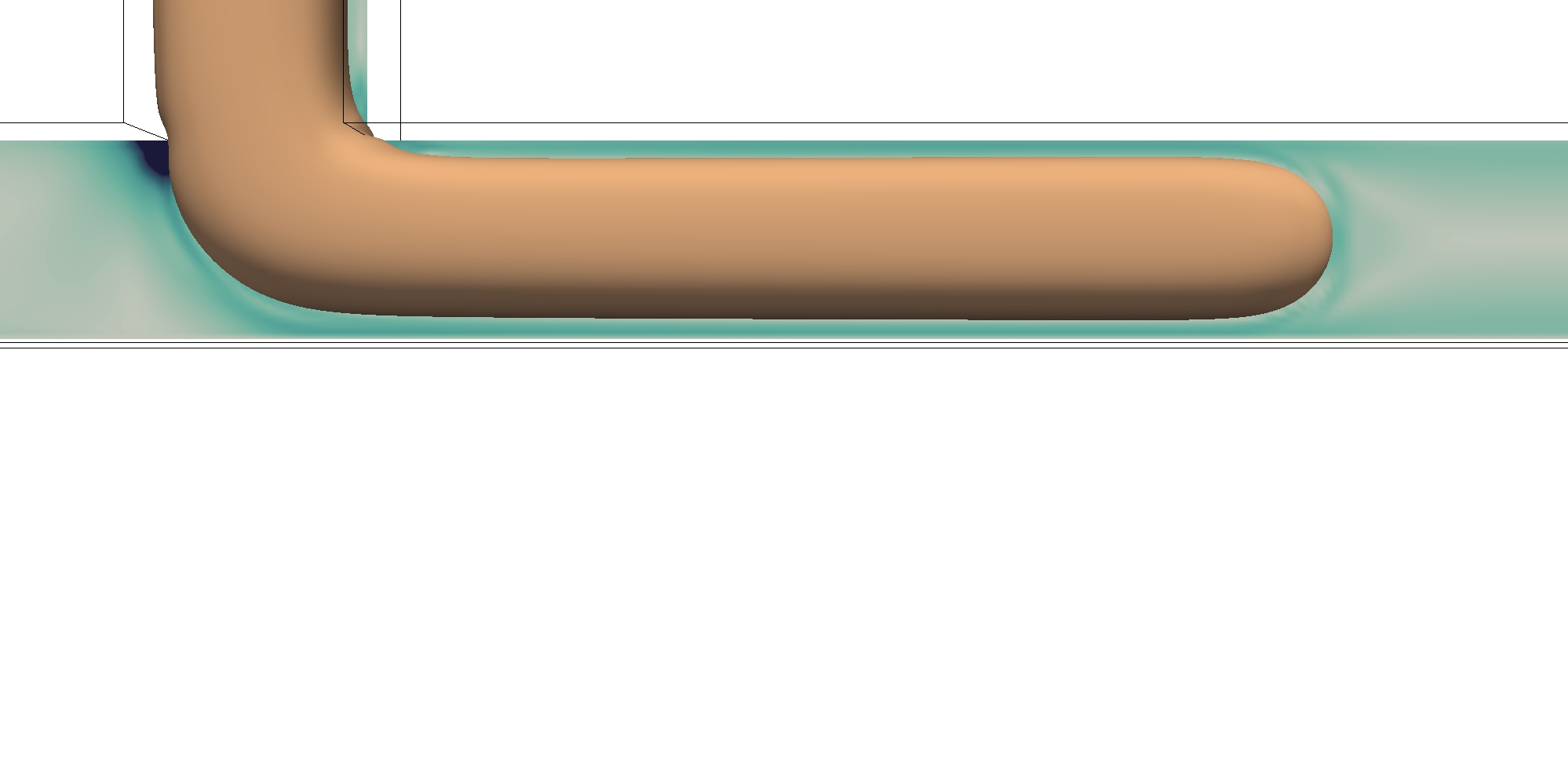}}\\
\subfigure[\,{\scriptsize $t = t_0 + 5.250 \, \tshear; \Ma = 1.26; \nabla T < 0$}]{\includegraphics[trim=10 390 0 0 , clip, width = 1.00\linewidth]{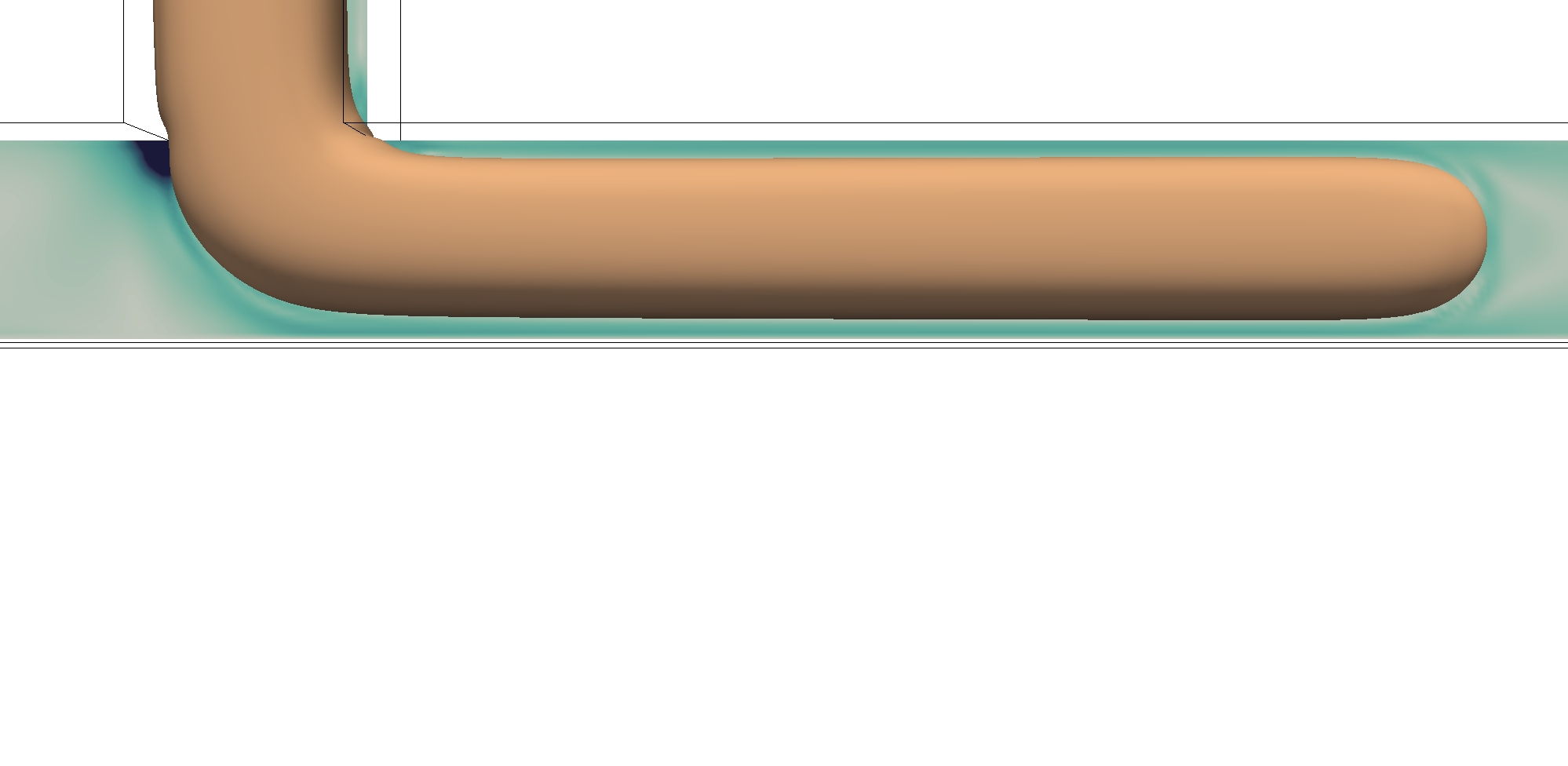}}
\end{minipage}
\caption{Numerical simulations in the confined T-junction geometry in presence of thermocapillarity. The importance of the thermocapillary stresses is quantified with the Marangoni number ($\Ma$, see \eqref{eq:Ma}) at changing also the temperature gradient $\nabla T$. Panels (a), (f), (k): temperature across the T-junction for three representative cases: $\Ma = 0$ (a); $\Ma = 1.26$, $\nabla T > 0$ (f); $\Ma = 1.26$, $\nabla T < 0$ (k). Low temperature regions appear as light (light green) while high temperature regions are dark (dark green). The other panels report density contours during the droplet formation process overlaid on the hydrodynamic viscous stress in the continuous phase: $\Ma = 0$ (Panels (b)-(e)); $\Ma = 1.26$, $\nabla T > 0$ (Panels (g)-(j)); $\Ma = 1.26$, $\nabla T < 0$ (Panels (l)-(o)). Low stress regions appear as light (light blue) while high stress regions are dark (blue). In all the cases the Capillary number ($\Ca$, see \eqref{eq:Ca}) is fixed to $\Ca = 0.001$ and the flow-rate ratio ($\phi$, see \eqref{eq:Q}) is fixed to $\phi = 1.0$. Notice that we have used the characteristic shear time $\tau_{\mbox{\tiny{shear}}}=H/\vc$ as a unit of time. \label{fig:Ca_001_rho1}}
\end{figure*}

We start our numerical investigation by getting a qualitative idea of the thermocapillary effects in the confined T-junction geometry. Specifically, in Fig.~\ref{fig:Ca_001_rho1} we present a time series of 3D snapshots of density contours overlaid on the viscous stress magnitude at fixed $\Ca$ for different scenarios, depending on the sign of the temperature gradient. The Marangoni number $\Ma=1.26$ is kept fixed. Moreover, we have chosen a relatively small $\Ca = 0.001$, which falls into the squeezing regime~\cite{Demenech07,LiuZhang09,LiuZhang11,BowerLee11}, where the incoming thread tends to occupy and obstruct the cross-section of the main channel before breaking at the junction. In this regime, the interfacial forces are dominant, and therefore we expect the thermocapillary effects to be more pronounced. Cases with zero temperature gradient - hence no thermocapillary effects - are reported in the left column (Panels (a)-(e)), and they are useful to compare with the cases where the temperature gradient is different from zero. In the middle column we report cases with a positive temperature gradient and $\Ma = 1.26$ (Panels (f)-(j)), while the right panel reports cases with negative temperature gradient and $\Ma = 1.26$ (Panels (k)-(n)). Notice that on the top row we have reported the temperature profile in a slice taken at $z=H/2$ for the three cases analyzed. Without thermocapillary effect, the thread starts to enter the main channel (Panel (b)), obstructs the main channel (Panel (c)), the continuous phase squeezes the dispersed phase (Panel (d)), and finally the droplet generation takes place at the junction (Panel (e)). The droplet formation process is visibly influenced by the switch-on of a non-zero temperature gradient: when $\Ma = 1.26$, $\nabla T > 0$ we observe that the droplet size is smaller than the corresponding case with $\nabla T = 0$; additionally, the break-up process is faster, leading to the production of two droplets within the same time it takes to generate one droplet with $\nabla T = 0$. On the contrary, when $\Ma = 1.26$, $\nabla T < 0$, the break-up process is delayed, and no droplet formation is observed within the time lapse of a few shear times $\tau_{\mbox{\tiny{shear}}}=H/\vc$, which is taken as the characteristic unit of time for the break-up process. The case with negative temperature gradient is also visibly characterized by a reduction of the viscous stress at the junction, which may be taken as a further indication that the whole process is evidently affected by thermocapillary stresses.\\

\begin{figure}[h!]
\begin{center}
\includegraphics[width = 0.47\linewidth]{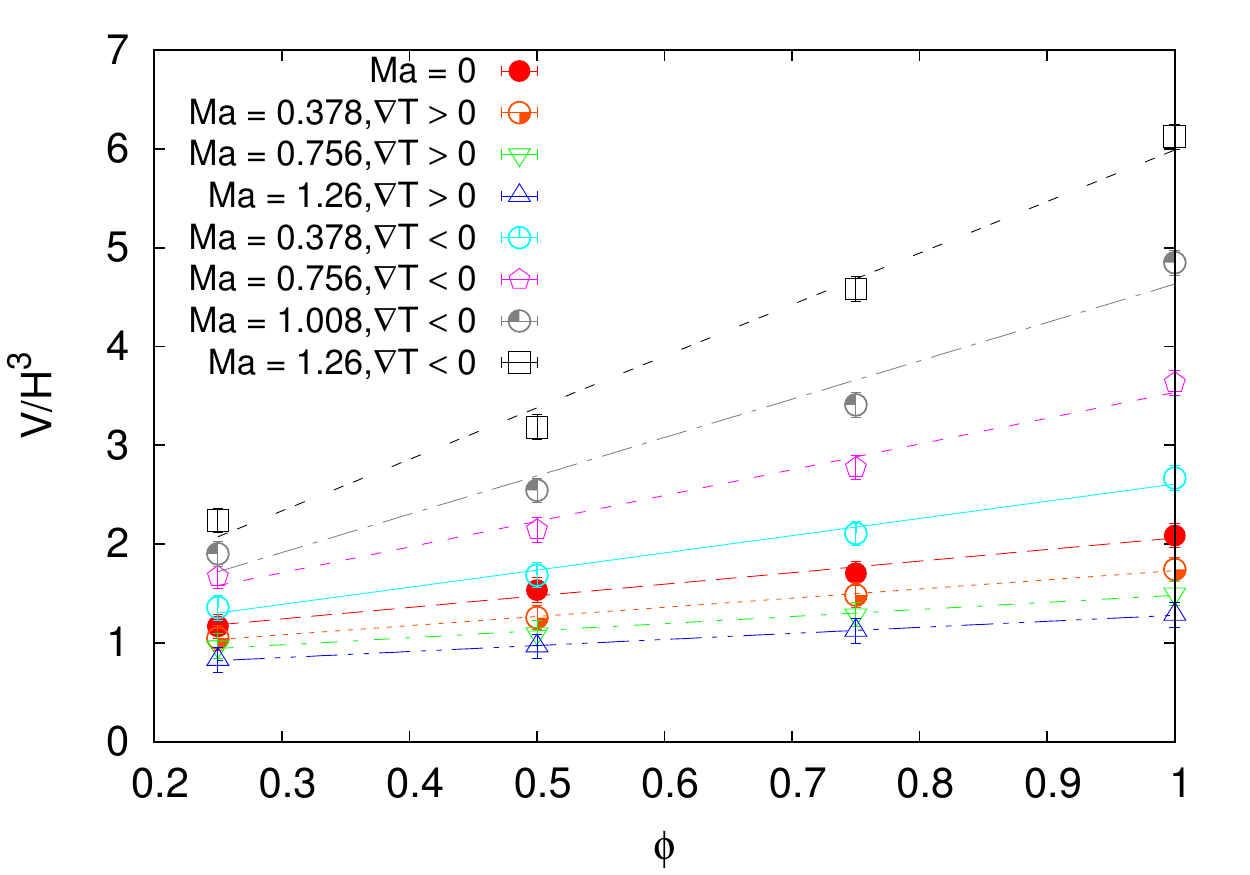}
\includegraphics[width = 0.47\linewidth]{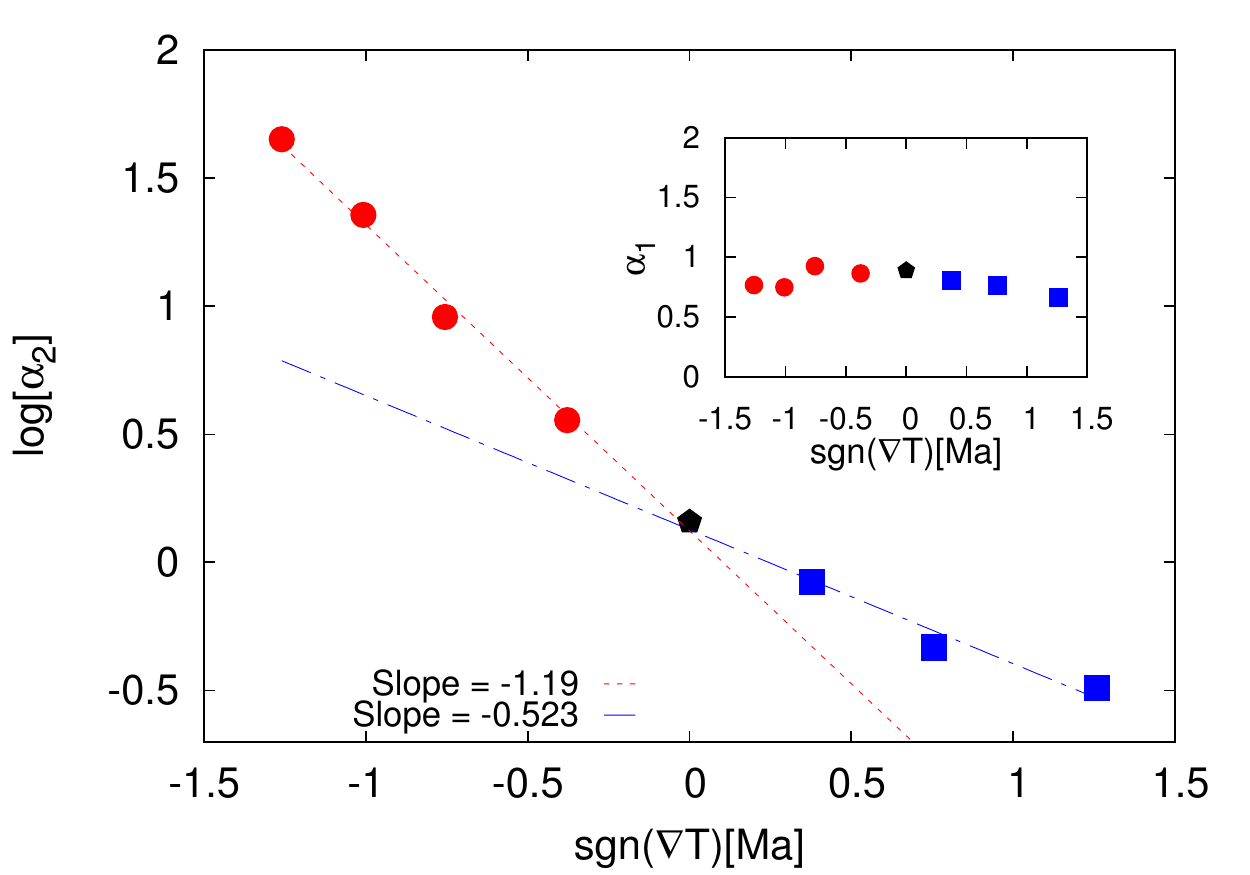}
\caption{Quantitative analysis for the dimensionless droplet volume as a function of the flow-rate ratio $\phi$ in the small Capillary number ($\Ca$) regime. Left Panel: we report the scaling laws for the dimensionless droplet volume versus the flow-rate ratio $\phi$ in the squeezing \cite{Demenech07,LiuZhang11} regime at $\Ca = 0.0014$, for different values of the Marangoni number ($\Ma$). Both negative and positive temperature gradients $\nabla T$ are considered. Linear fits are drawn to guide the eyes. Right Panel: analysis of the linear scaling law~\eqref{eq:Garstecki} for the data in the left Panel. We report the scaling factor $\alpha_2$ versus $\sgn(\nabla T)[\Ma]$. We also provide two slope fits for the data at $\nabla T \le 0$ (red circles) and $\nabla T \ge 0$ (blue squares). The inset shows the plot of $\alpha_1$ versus $\sgn(\nabla T)[\Ma]$. \label{fig:scaling_thermo}}
\end{center}
\end{figure}

\begin{figure}[h!]
\begin{center}
\includegraphics[width = 0.47\linewidth]{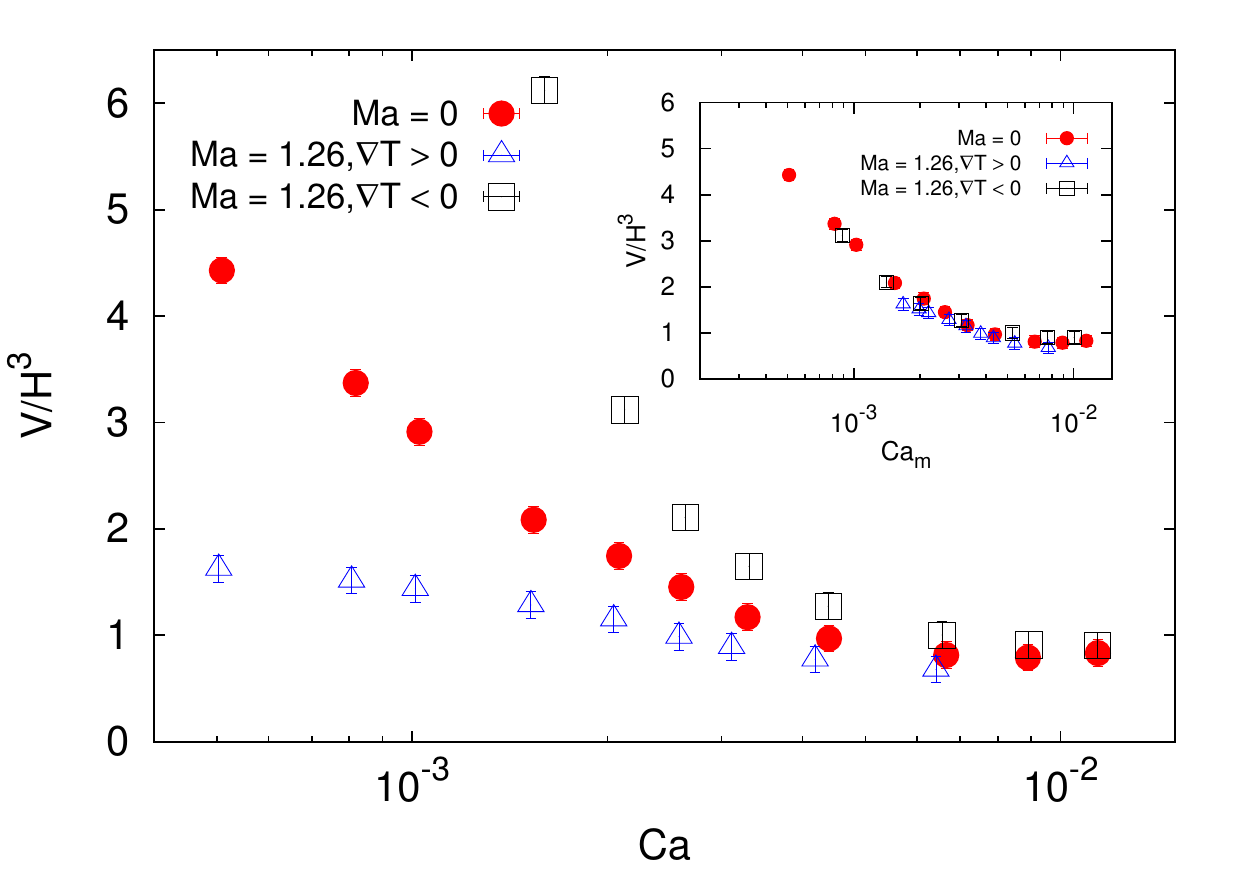}
\includegraphics[width = 0.47\linewidth]{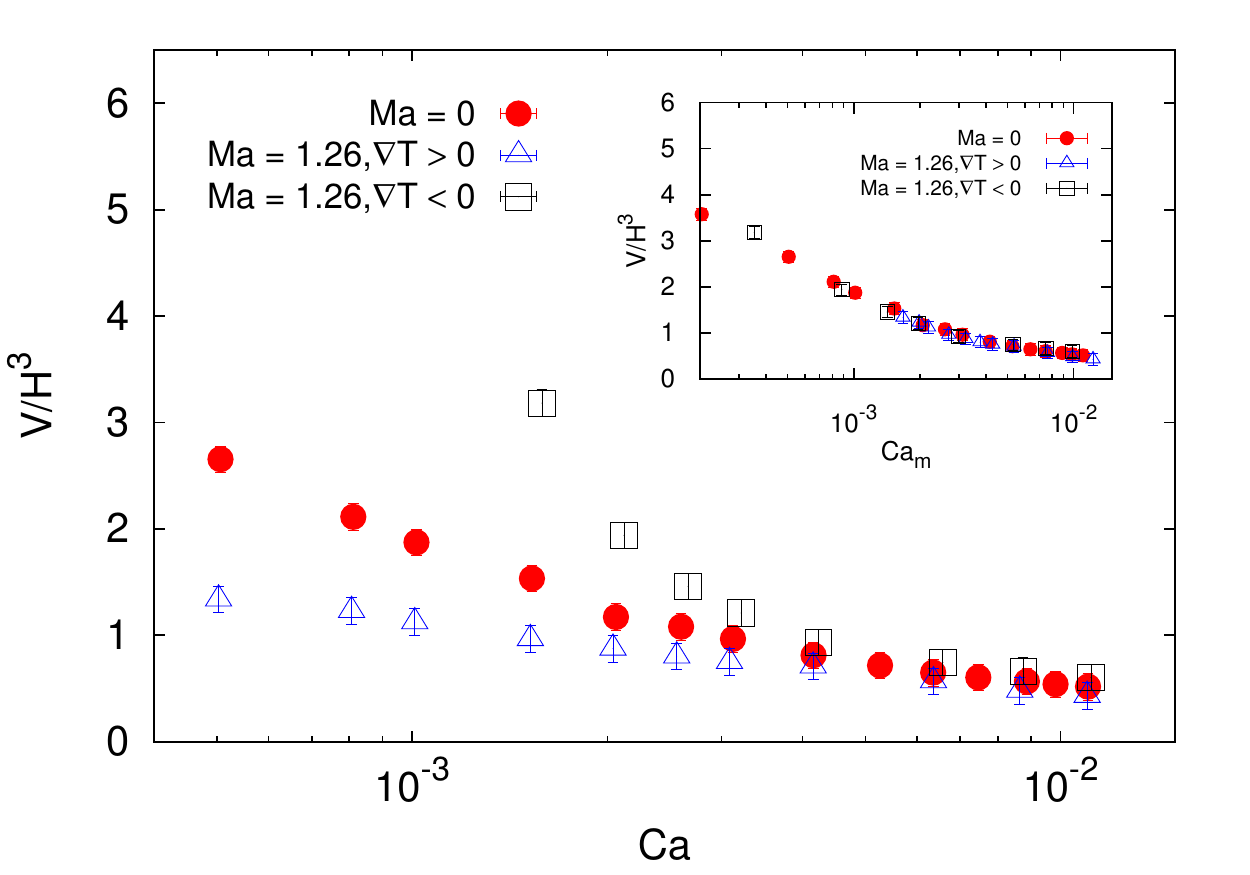}
\caption{Quantitative analysis for the dimensionless droplet volume as a function of the Capillary number ($\Ca$), for flow-rate ratio $\phi = 1.0$ (left Panel) and $\phi = 0.5$ (right Panel). For the Marangoni number ($\Ma$), we have chosen the reference case $\Ma = 0$ (red circles), and the value $\Ma = 1.26$ for both negative (black squares) and positive (blue triangles) temperature gradients $\nabla T$. The two insets show the plot of the dimensionless droplet volume versus the rescaled Capillary number ($\Cam$) defined in~\eqref{eq:scaled_Ca}, where a characteristic velocity $\vm$ is introduced to account for the thermocapillary effects. \label{fig:droplet_size}}
\end{center}
\end{figure}

\begin{figure}[h!]
\begin{center}
\includegraphics[width = 0.47\linewidth]{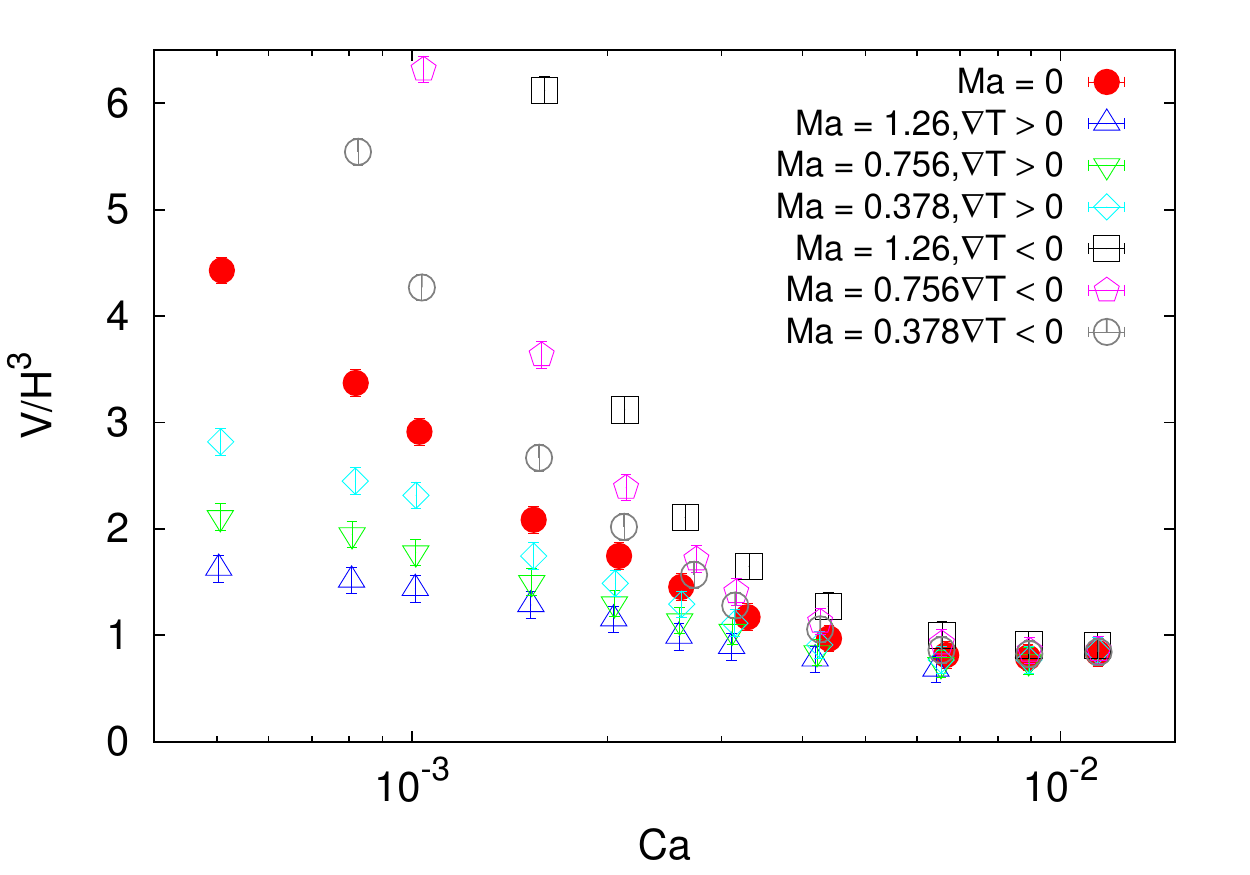}
\includegraphics[width = 0.47\linewidth]{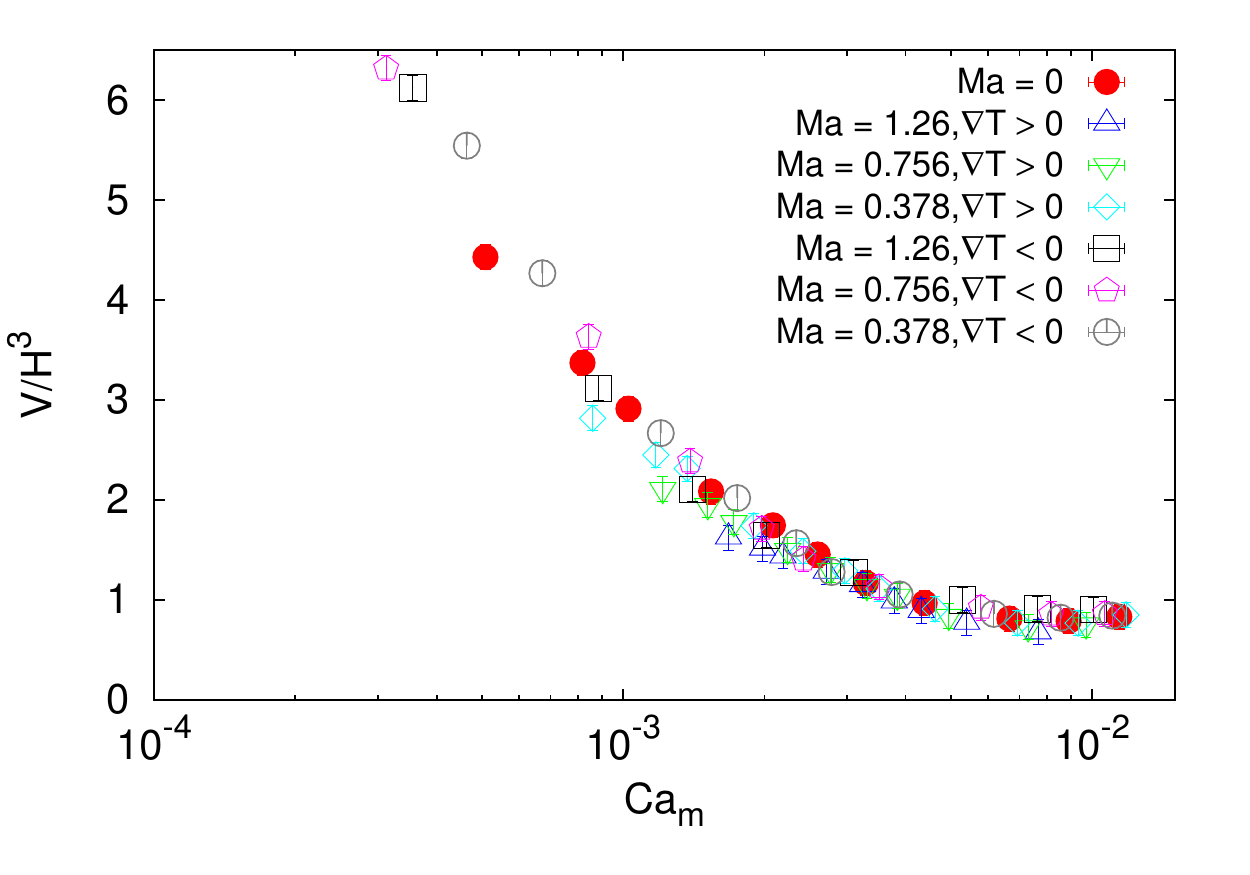}
\caption{Quantitative analysis for the dimensionless droplet volume as a function of the Capillary number ($\Ca$), for different values of the Marangoni number ($\Ma$). Both negative and positive temperature gradients $\nabla T$ are considered. Left Panel: we report the dimensionless droplet volume versus $\Ca$ for different values of $\Ma$ and $\nabla T$. In all cases, the flow-rate ratio is kept fixed to $\phi = 1.0$. Right Panel: data in the left Panel are plotted against the rescaled Capillary number ($\Cam$) defined in~\eqref{eq:scaled_Ca}. Notice that the value of the thermocapillary velocity $\vm$ is changing with $\Ma$, and is determined by simply rescaling the value of $\vm$ found for $\Ma=1.26$ in Fig.~\ref{fig:droplet_size}. \label{fig:thermo_branch}}
\end{center}
\end{figure}

To go deeper and characterize the effects of thermocapillarity on the droplet size at small $\Ca$, we have further performed different numerical simulations at changing the flow-rate ratio, the Marangoni number, and the sign of the temperature gradient. This is necessary to compare our results for the droplet size with the scaling laws which have been proposed in the literature~\cite{Demenech07,LiuZhang11}. As already highlighted in Fig.~\ref{fig:Ca_001_rho1}, in the regime of small $\Ca$, the dispersed phase enters the main channel until it obstructs the flow. Afterwards, due to the squeezing of the continuous phase, the dispersed thread breaks at the T-junction~\cite{Demenech07,LiuZhang09}. Assuming that the thread squeezes at a rate proportional to $Q_c$, a scaling law for the droplet volume has been established in the literature~\cite{Garstecki06,Demenech07}
\begin{equation}\label{eq:Garstecki}
\frac{V}{H^3}= \alpha_1 + \alpha_2 \phi .
\end{equation}
For a fixed Capillary number, the constants $\alpha_1$ and $\alpha_2$ (which are of the order one) are determined by the geometrical details of the channels~\cite{LiuZhang11,Demenech07}. As already put forward by other authors~\cite{Demenech07,Garstecki06}, the relation \eqref{eq:Garstecki} allows to match together two scaling laws expected in the regimes where $\Qd$ is smaller or larger than $\Qc$. When $\Qd$ is larger than $\Qc$, one expects that the time to squeeze the droplet is larger than the time it takes to block the channel, which implies the droplet size to scale linearly with the flow-rate ratio. On the other hand, when $\Qd$ is much smaller than $\Qc$, one expects the dimensionless droplet size to be independent of the flow-rate ratio~\cite{Demenech07,Garstecki06}. Equation~\eqref{eq:Garstecki} has already been verified in experiments~\cite{Garstecki06,Christopher08,Glawdeletal} and also in numerical simulations~\cite{Demenech07,LiuZhang09,LiuZhang11,BowerLee11,gupta2016effects,gupta2016lattice}. It is therefore important to check both the prediction of a linear scaling in $\phi$, as well as a possible dependency of the scaling constants $\alpha_1$ and $\alpha_2$ on the Marangoni number and the sign of the temperature gradient. To this aim, in the left Panel of Fig.~\ref{fig:scaling_thermo}, we plot the dimensionless droplet volume versus the flow-rate ratio at $\Ca = 0.0014$, for different values of $\Ma$ and different signs of the temperature gradient. For all the cases investigated, we confirm the validity of the linear scaling relation~\eqref{eq:Garstecki}, with the curves that are steeper for negative temperature gradients. Notice that the effect of thermocapillarity is larger (smaller) at the larger (smaller) $\phi$, which may be taken as a clear indication that one needs a sufficiently large squeezing time in order for the Marangoni stresses to show-up with a quantitative effect. To make further progress, from the data reported in the left Panel of Fig.~\ref{fig:scaling_thermo}, we have obtained the constants $\alpha_1$ and $\alpha_2$ from the linear fit based on Eq.~\eqref{eq:Garstecki}. These two coefficients are displayed against the Marangoni number in the right Panel of Fig.~\ref{fig:scaling_thermo}. Notice that, in order to visualize together the effects of both positive and negative temperature gradients, we use  $\sgn \nabla T [\Ma]$ in the horizontal axis. The constant $\alpha_1 \approx {\cal O}(1)$ is not appreciably affected by the thermocapillary effects. The effect is actually more visible on the slope coefficient $\alpha_2$. The plot of $\log \alpha_2$ indeed shows two distinguished slopes, dependently on the sign of the temperature gradient. This change in slope is somehow indicative that the geometry mediated break-up process is affected in a non trivial way by the thermocapillary effects: for $\nabla T < 0$ the thermocapillary effects oppose to the squeezing process, producing larger droplet sizes and somehow ``pushing'' the system even more towards the squeezing regime, favoring the inhibition of the main channel by the dispersed phase (see also right column in Fig.~\ref{fig:Ca_001_rho1}). When $\nabla T > 0$, instead, the thermocapillary stresses promote droplet break-up and somehow spoil distinctive features of the squeezing regime, as also evident from the flatter curves displayed in the left Panel of Fig.~\ref{fig:scaling_thermo}. The latter observation stimulates another interesting analysis, concerning the importance of the thermocapillary effects in the transition from the squeezing regime to the dripping regime~\cite{Demenech07,LiuZhang09,LiuZhang11,BowerLee11,gupta2016effects,gupta2016lattice}, where the increase of the Capillary number induces a reduced droplet size. For the specific case of our numerical simulations, this analysis is displayed in Fig.~\ref{fig:droplet_size}, where we monitor the droplet size by changing the Capillary number for fixed flow-rate ratio. We considered two cases with flow-rate ratio $\phi =1.0$ (left Panel) and $\phi = 0.5$ (right Panel). Furthermore, the effects of thermocapillarity are also shown, in that we have conducted simulations by switching-on a positive (blue triangles) and a negative (black squares) temperature gradient for the same Marangoni number $\Ma=1.26$. For a given $\Ma$ and $\nabla T$, the droplet size decreases as a function of $\Ca$, due to the increased importance of the viscous stresses over the surface tension forces. We actually see that for the largest $\Ca$ analyzed the droplet sizes come out to be comparable for all the cases, while at smaller $\Ca$ data with a positive (negative) temperature gradient underestimate (overestimate) the ``bare'' data, i.e. those  with zero temperature gradient. These facts said, we have tested the possibility to design a quantitative guideline to predict the droplet size, once the bare Newtonian data and the input thermocapillary parameters are known. Specifically, we checked whether a suitable redefinition of the Capillary number in terms of the input thermocapillary parameters allows to rescale the results with positive/negative temperature gradients on the bare Newtonian data. Notice that, once the temperature gradient and the Marangoni number are assigned, a characteristic Marangoni velocity $\vm$ naturally follows, which scales linearly with $\Ma$. With such velocity at hands, one can define a rescaled Capillary number as 
\be \label{eq:scaled_Ca}
\Cam = \frac{\etac (\vc+\vm)}{\sigma}.
\ee
Equation~\eqref{eq:scaled_Ca} takes into account the balance between the viscous forces and surface tension forces, including also a modification coming from the Marangoni velocity $\vm$, whose sign is positive/negative for positive/negative temperature gradients. We have then considered the data in the main Panels of Fig.~\ref{fig:droplet_size} and tested this rescaling. Results are reported in the insets of Fig.~\ref{fig:droplet_size}. Remarkably, by using a unique value of the Marangoni velocity $\vm=\pm 0.00035$ lbu (for $\Ma = 1.26$), we are able to rescale the thermocapillary data on the top of the bare Newtonian data. This holds true for both the flow-rate ratios $\phi = 1.0$ and $\phi = 0.5$. We emphasize that we have not predicted the value of the thermocapillary velocity $\vm$, but just found it at a fixed $\Ma$ on the basis of the best collapse of the data. Nevertheless, we notice that the value falls well in the ballpark of the values of steady migration velocities that we have measured and reported in the benchmark tests described in Appendix~\ref{sec:benchmark}. We further checked if the rescaling~\eqref{eq:scaled_Ca} performs well at changing the Marangoni number. In Fig.~\ref{fig:thermo_branch} we report the droplet size as a function of $\Ca$ for $\phi = 1.0$; different values of the Marangoni numbers are considered, for both positive and negative temperature gradients. The rescaling velocity $\vm$ is now determined a priori, by a simple linear rescaling of the characteristic velocity $\vm$ found for $\Ma=1.26$, i.e. $\vm=\vm^{(0)} \, \Ma/1.26$ ($\vm^{(0)}=\pm 0.00035$ lbu), hence no extra fitting parameters have been introduced. As we can see from the figure, all the data nicely collapse on a single master curve, which is a clear indication that the proposed argument is robust at changing the Marangoni number, at least up to $\Ma \approx 1$.\\ 
Finally, some points of discussion emerge, to connect the results of the present numerical simulations to experimental data, especially with the aim of proposing a set of experimental suggestions. Experimental literature indeed exists (see~\cite{selva2011temperature,Baroud07,Yap09,Delville12} and references therein), reporting on the motion of droplets and threads in confined channels with thermocapillary forces. For example, the authors in~\cite{selva2011temperature} report on the experimental study of temperature-induced migration of a bubble in a Hele-Shaw cell. In the experiments, the bubble velocity hits a maximum value $\simeq 800 \mu$ m/s, and it depends on the temperature gradient ($\simeq$ power supply of $65-300$ mW), bubble radius ($20-400 \mu$m) and the wall thickness ($22$ and $37 \mu$m). Thus, in an experiment involving the motion of threads in microfluidic T-junctions with characteristic flow-rates of the order of $Q_{\rm c,d} \approx 0.1 \mu $L/min~\cite{Garstecki06,Micromachines} and channel lengths of the order of a few tens of micrometers (say $H \approx W \approx 100 \mu$m), one expects velocities of the order of $v_{\rm c,d} \simeq 100 \mu$m/s, thus the thermocapillary effects are expected to show-up in sizeable effects.\\

\section{Conclusions}\label{sec:conclusions}

One of the main features that drives the interest in developing and exploiting microfluidic devices is the possibility to control the process of droplet formation as a function of the flow and material properties~\cite{Christopher07,Seeman12,Christopher08,Teh08,Baroud10}. The complex flows in these devices are naturally accompanied by confinement, which has an important effect on the droplet break-up process. In this context, droplet manipulation based on thermocapillary stresses has been demonstrated both experimentally and numerically (see~\cite{Baroud07,Yap09,Delville12,Zhangetal12,Zhangetal13,Zhangetal14,Li15} and references therein). Thermocapillary stresses arise along an interface because of a temperature dependency of the surface tension. Based on lattice Boltzmann simulations with the ``Shan-Chen'' model for non-ideal binary mixtures~\cite{SC93,SC94,Shan08,CHEM09,SbragagliaBelardinelli}, we have investigated the combined effects of thermocapillarity and confinement on the dynamics and break-up of fluid threads in a microfluidic T-junction geometry, where a dispersed phase is injected perpendicularly into a main channel containing a continuous flowing phase which induces periodic formation of droplets. The numerical model for multicomponent fluids~\cite{SC94,SbragagliaBelardinelli} has been first extended to include the effects of thermocapillarity, and then has been applied to the simulations in the confined T-junction geometry, by focusing the attention on the operational regimes of ``squeezing'' and ``dripping'' \cite{Demenech07} with moderate flow-rate ratios. We have changed various parameters, including the temperature gradient $\nabla T$, the Capillary number $\Ca$, and the Marangoni number $\Ma$, to get a quantitative understanding on how the thread dynamics and subsequent break-up are affected by the thermocapillary forces. To this aim, we wish to stress the crucial role played by numerical simulations, in allowing a systematic exploration of the importance of the relevant free parameters in the model. With the data at $\Ma=0$ at hand (the ``bare'' data), we have indeed tested the possibility to design guidelines to predict the droplet size for the cases where thermocapillary effects are present ($\Ma \neq 0$). Numerical results indicate that a suitable redefinition of the Capillary number based on a characteristic thermocapillary velocity well accounts for the observed changes and allows to well rescale the thermocapillary data on the top of the bare data. Remarkably, this simple argument is robust at changing both the flow-rate ratio and the Marangoni number, up to $\Ma \approx 1$. Quantifying such rescaling properties at higher $\Ma$ sets interesting challenges for the future: for the description of the thermocapillary migration of a droplet at very large $\Ma$, previous studies suggest non trivial dependencies on $\Ma$  (see \cite{Zhangetal12} and references therein); our model is able to simulate flows at those Marangoni numbers (see Appendix~\ref{sec:benchmark}), and therefore it is definitively warranted a systematic investigation, to verify if the aforementioned rescaling holds and if the rescaling velocity is still linearly dependent on $\Ma$ or not \cite{Zhangetal12}. Along these lines, it is also definitively warranted a systematic exploration with fluids with dissimilar properties (viscosity ratio and/or thermal conductivities) as well as comparisons with other existing LB methodologies for thermocapillary flows~\cite{Zhangetal12,Zhangetal13,Zhangetal14,Li15}. We actually expect our numerical work can stimulate further investigation and can also bring forth experimental suggestions for droplet control in microfluidic geometries.\\

We kindly acknowledge funding from the European Research Council under the Europeans Community's Seventh Framework Programme (FP7/2007-2013) / ERC Grant Agreement N. 279004. We acknowledge the computing hours from ISCRA B project (COMPDROP), CINECA Italy. The authors acknowledge useful discussions with Dr. H. Liu. 

\appendix

\section{Benchmark Tests: Droplet migration}\label{sec:benchmark}

\begin{figure}[t!]
\begin{center}
\includegraphics[scale=0.7]{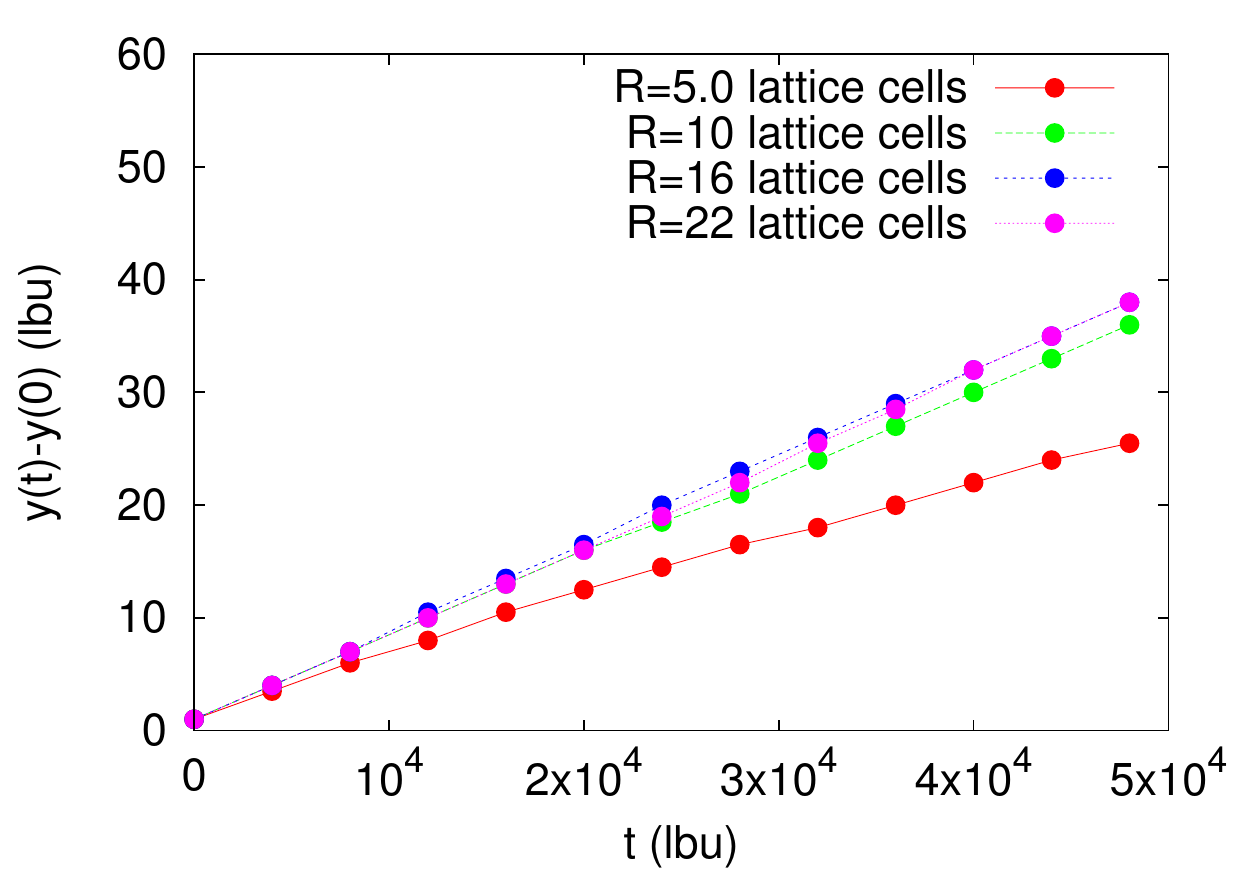}
\caption{We report the evolution in time of the center of mass of a droplet (relative to its initial value) steadily translating in a temperature gradient. The droplet radius $R$ is resolved with a variable number of lattice cells, and the temperature gradient is consequently adjusted to match the same Marangoni number $\Ma=1.25$. Other simulation details are described in the text. \label{fig:resolution}}
\end{center}
\end{figure}

\begin{figure}[h!]
\begin{center}
\includegraphics[scale=0.7]{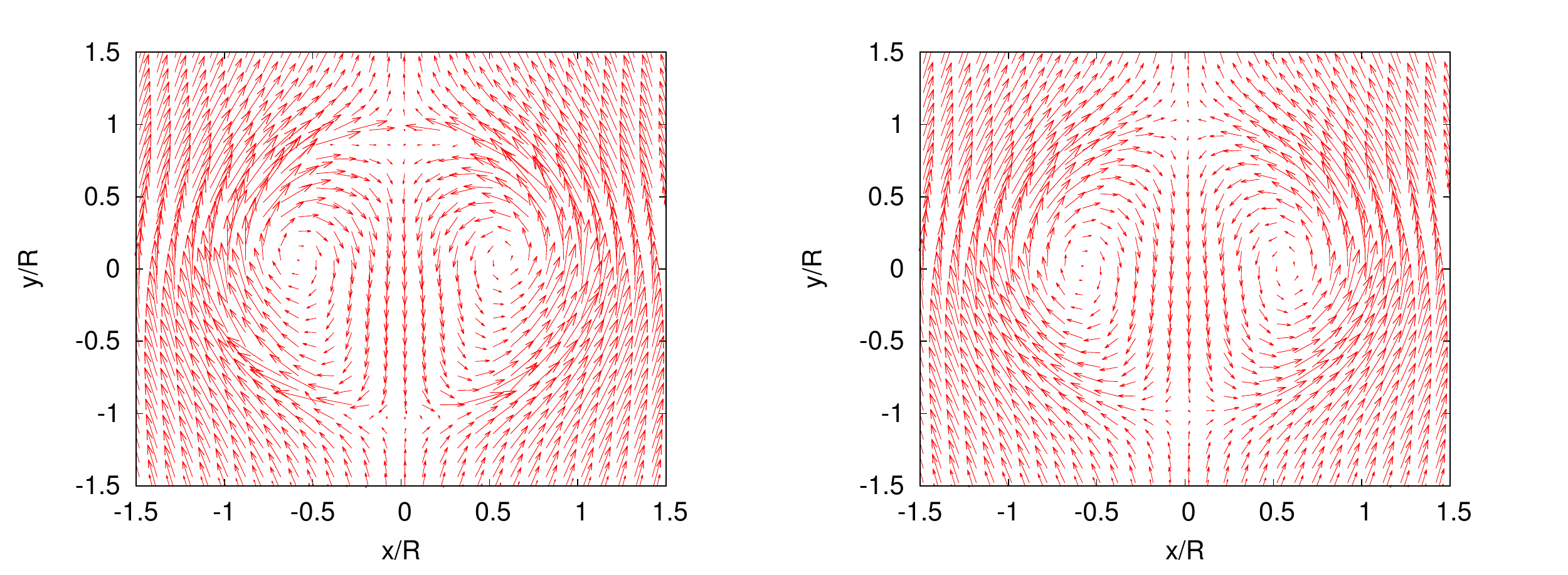}
\caption{The velocity profiles in the reference frame of the moving droplet steadily translating in a temperature gradient. The droplet moves from top to bottom. Results from numerical simulations at $\Ma=5.0$ (left Panel) are compared with the analytical profiles (right Panel, See Appendix \ref{sec:Appendix}).\label{fig:1}}
\end{center}
\end{figure}

\begin{figure}[h!]
\begin{center}
\includegraphics[scale=0.7]{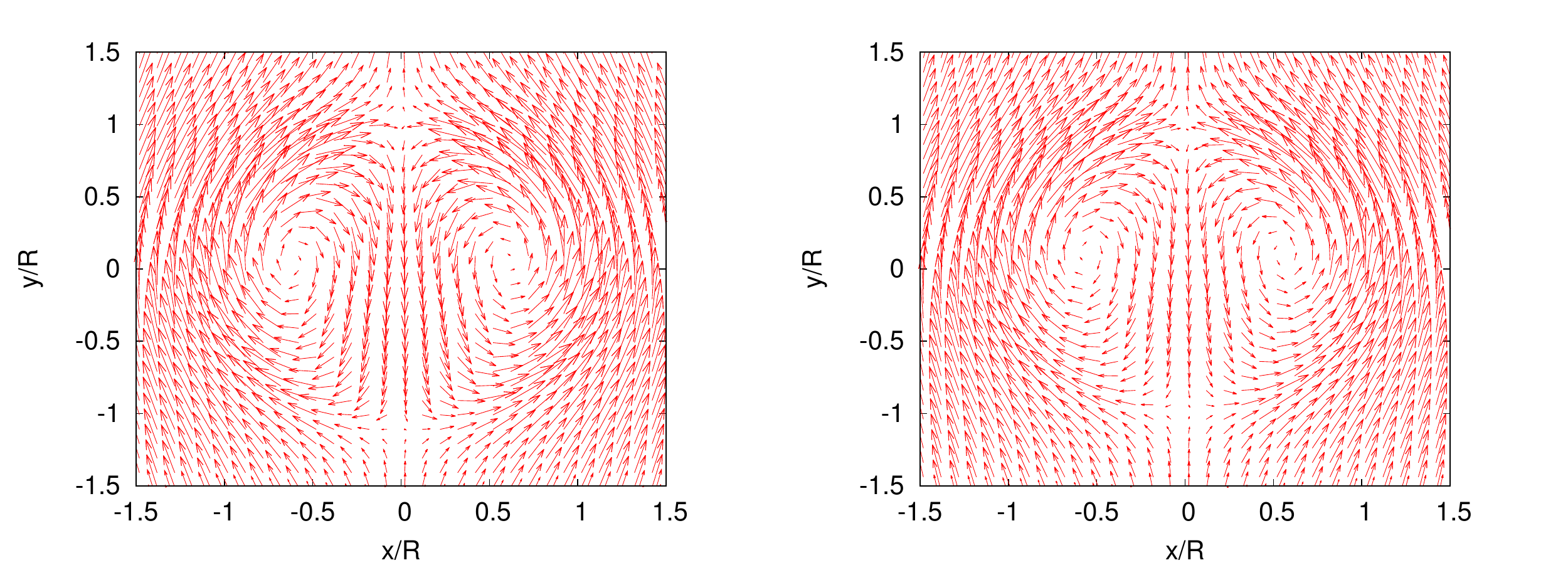}
\caption{The velocity profiles in the reference frame of the moving droplet steadily translating in a temperature gradient at changing the Marangoni number $\Ma$: $\Ma=20$ (left Panel); $\Ma=50$ (right Panel). \label{fig:1b}}
\end{center}
\end{figure}

\begin{figure}[h!]
\begin{center}
\includegraphics[scale=0.7]{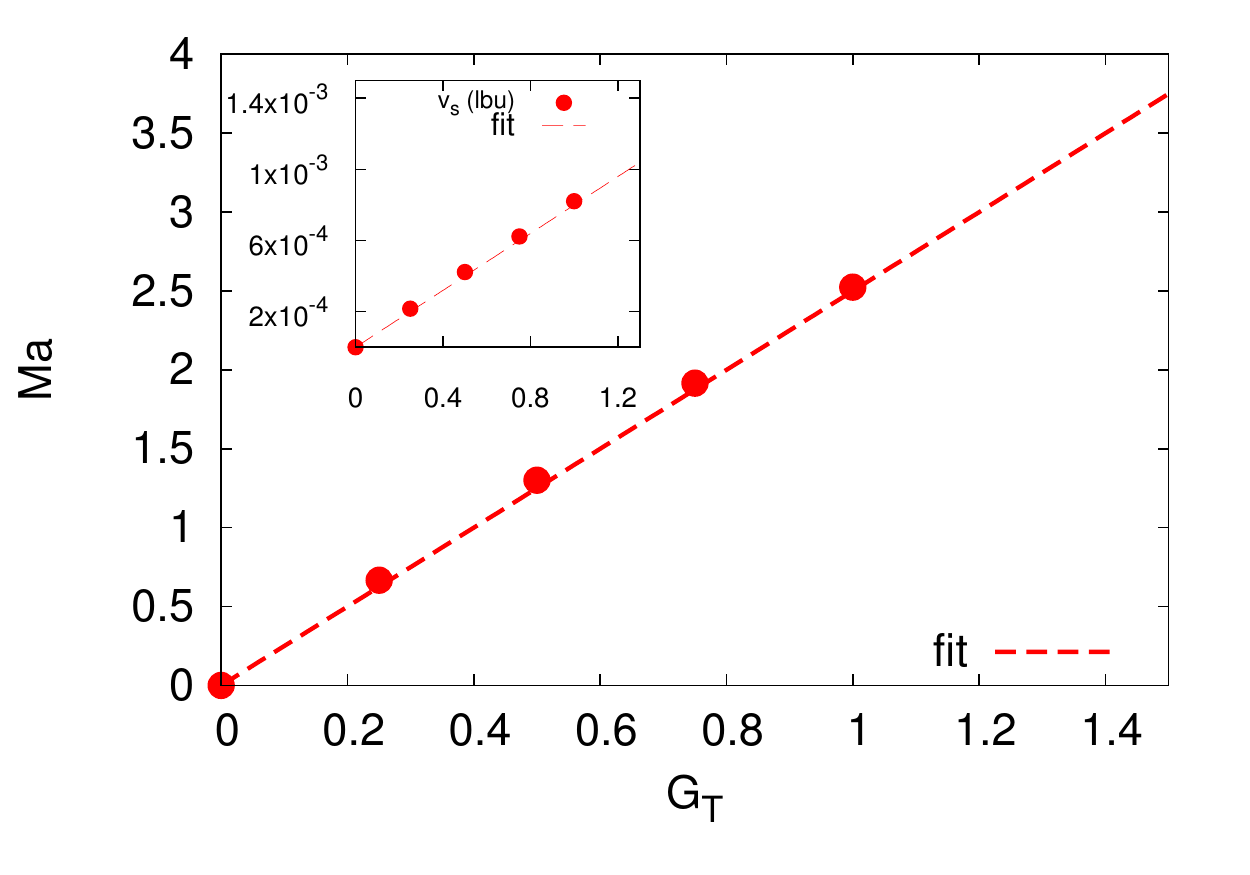}
\includegraphics[scale=0.7]{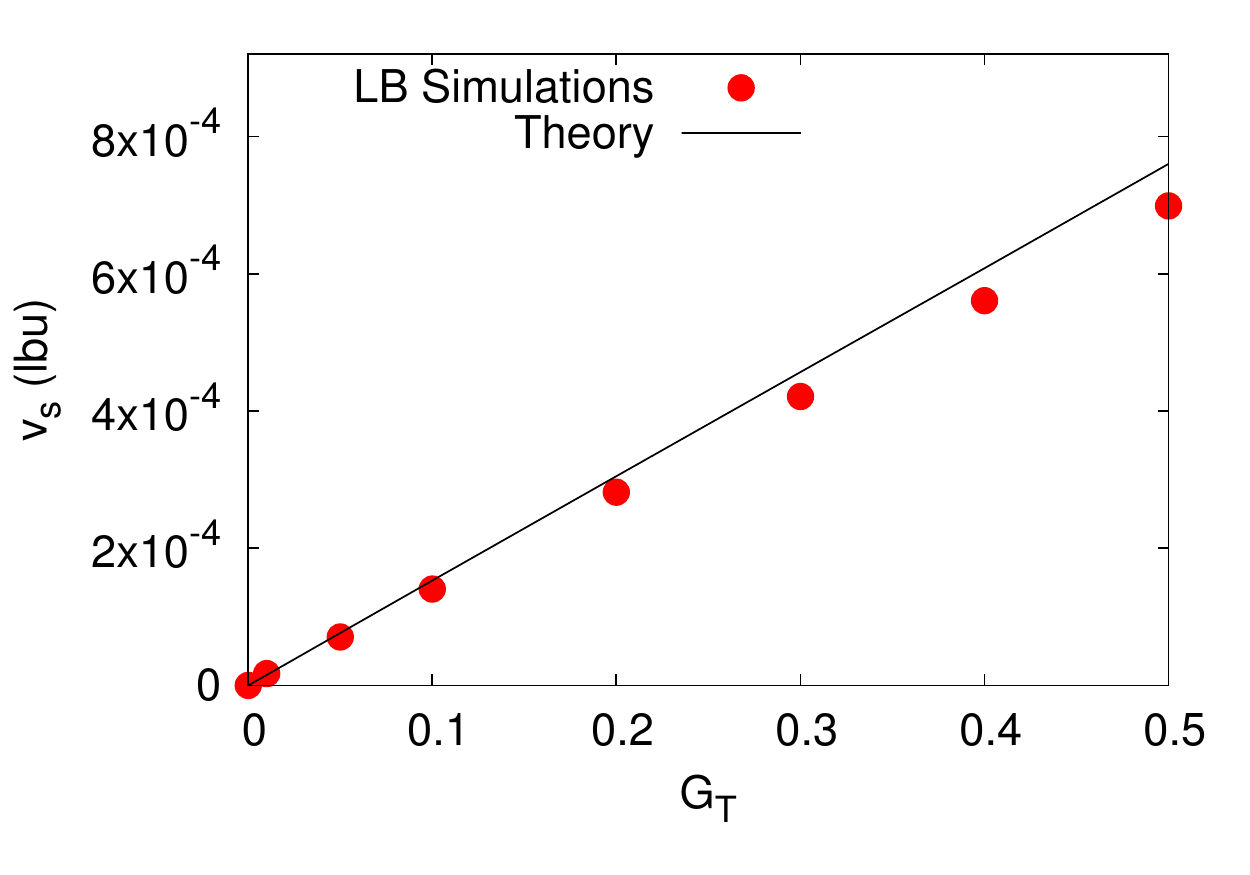}
\caption{Left Panel: we report the Marangoni number ($\Ma$) \eqref{eq:Ma} as a function of the parameter $\GT$~\eqref{EFFEKAPPA}. The Marangoni number is computed from the steady velocity of migration of a droplet in a constant temperature gradient described in section~\ref{sec:benchmark}. Inset: we report the steady velocity~\eqref{terminalbistxt}. Right Panel: the black line stands for the rhs of Eq.~\eqref{terminalbistxt}, with the jump in surface tension $\Delta \sigma_{2R}$ estimated from the modified pressure tensor, as described in Section~\ref{sec:LBmulti}. The points represent the steady velocity of migration obtained from the numerical simulations. \label{fig:2}} 
\end{center}
\end{figure}

\begin{figure}[h!]
\begin{center}
\includegraphics[scale=0.7]{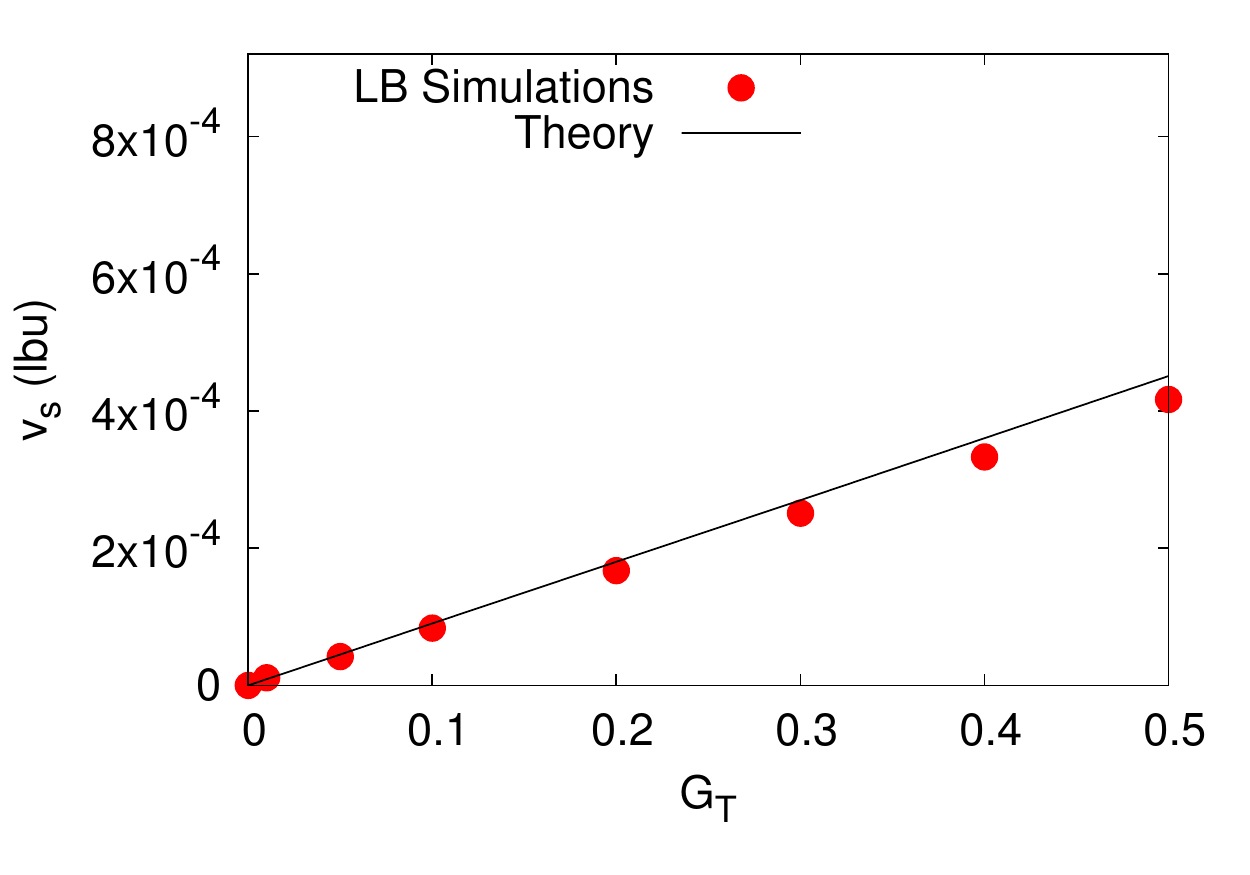}
\includegraphics[scale=0.7]{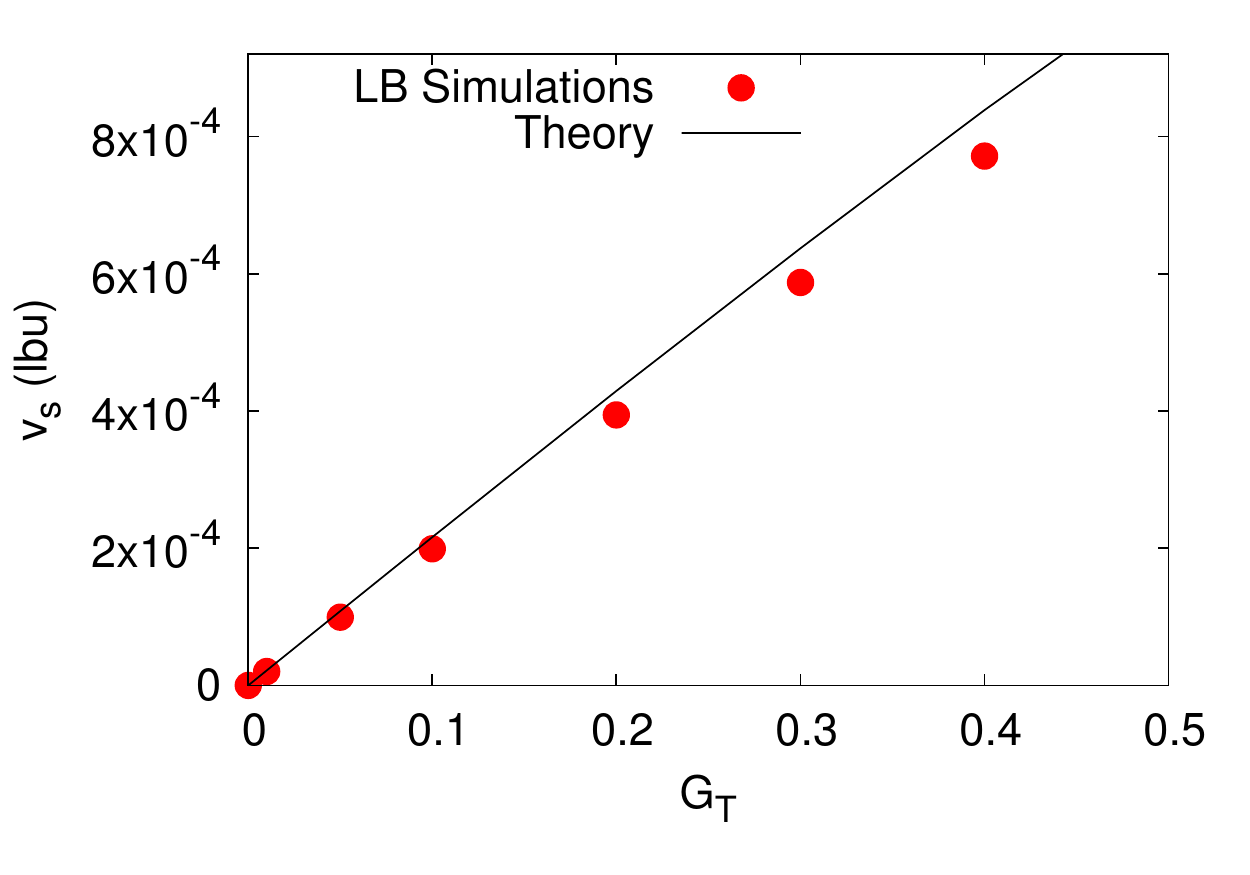}
\caption{The analysis shown in the right Panel of Fig.~\ref{fig:2} is extended to fluids with dissimilar thermal conductivities: $\kappac^{(T)}=0.06$, $\kappad^{(T)}=0.166$ lbu (left Panel) and $\kappad^{(T)}=0.06$, $\kappac^{(T)}=0.166$ lbu (right Panel). \label{fig:3}} 
\end{center}
\end{figure}

To perform benchmark numerical simulations for the thermocapillary migration, we choose a two dimensional (2D) set-up. Results are available from the literature~\cite{Subramanian01,Subramanianpapers1,Subramanianpapers2,Young59,Brady11}, concerning the motion of a droplet with radius $R$ in steady flows driven by thermocapillary effects in three dimensions (3D). As for the 2D case, the corresponding solution is straightforward to obtain, and details are reported in the Appendix~\ref{sec:Appendix}. The choice of a 2D set-up was initially done to easily approach data at larger resolution, to appreciate the convergence of the numerical results to the continuum limit predictions. Specifically, one can solve the problem of thermocapillary motion of a cylindrical droplet subject to a uniform temperature gradient in the absence of gravity. Assuming negligible advective effects on the momentum and heat transports, one can determine the constant speed $\vs$ of translation in the direction of the temperature gradient
\be\label{terminalbistxt}
\vs=\frac{\Delta \sigma_{2R}}{8 (\etad+\etac)}.
\ee 
In the above equation, $\Delta \sigma_{2R}=\Delta \sigma_{2R}(\kappac^{(T)},\kappad^{(T)})$ represents the variation in the surface tension between the rear and the front of the moving droplet, and the dependence from $\kappac^{(T)}$, $\kappad^{(T)}$ comes from the fact that the temperature at the interface is fixed by the ratio of the thermal conductivities (see Appendix~\ref{sec:Appendix} for details). Notice that for the case of a droplet, the Marangoni number is defined as
\be\label{eq:Ma2}
\Ma=\frac{2R \Delta \sigma_{2R} }{\kappac^{(T)} \etac}.
\ee
We performed numerical simulations in a 2D box with size $L_x \times L_y = 200  \times 640 $ lattice cells, with two walls located at $y=0$ and $y=L_y$. Periodic boundary conditions are applied in the $x$ direction. We performed different numerical tests, at varying the droplet size and the temperature gradient. To start with, in figure \ref{fig:resolution} we report the evolution in time of the center of mass of a droplet (relative to its initial value) steadily translating in a constant temperature gradient. The droplet radius is resolved with a variable number of lattice cells, and the temperature gradient is consequently adjusted to match the same Marangoni number, $\Ma=1.25$. As we can see, above $R \approx 15$ lattice cells, results show convergence. These converged results are then compared against their theoetical predictions. Specifically, from now on, we report details on a case with droplet radius $R=16$ lattice cells placed in the middle of the box, while the temperature of the upper and lower walls are fixed to be $T_{\mbox{\tiny{up}}}=-0.7$ lbu and $T_{\mbox{\tiny{down}}}=0.7$ lbu, respectively. We first consider a case with equal thermal conductivities inside and outside the droplet, $\kappac^{(T)}=\kappad^{(T)}=0.166$ lbu and same viscosity $\etac = \etad = 0.28$ lbu. The tipical velocity vector profiles in the reference frame of the moving droplet are shown in Fig.~\ref{fig:1}. The velocity vectors reveal recirculations, and the flow pattern is also compared with the analytical solution (right Panel) described in Appendix~\ref{sec:Appendix}. Although not used in the numerical simulations with the T-Junctions, the model is capable of handling also higher Marangoni numbers. Some velocity vectors profiles are reported in Fig.~\ref{fig:1b}. Further quantitative check is offered in Fig.~\ref{fig:2}, where we analyze the steady velocity $\vs$ as a function of the parameter $\GT$ in~\eqref{EFFEKAPPA}. Specifically, the constant $\GT$ in~\eqref{EFFEKAPPA} has been changed from $0.0$ to $1.0$. The inset of the left Panel reports the steady velocity (in lbu) that is developed due to the thermocapillary migration at changing $\GT$. Notice that for the set of parameters chosen, the Marangoni number  - hence and the steady velocity - are well fitted with a linear behaviour, $\Ma \approx 2.52 \, \GT$ ($\vs \approx 0.0008 \, \GT$). The right Panel reports a direct check of Eq.~\eqref{terminalbistxt}: the black line stands for the rhs of Eq.~\eqref{terminalbistxt}, with the jump in surface tension $\Delta \sigma_{2R}$ estimated from the modified pressure tensor described in Section~\ref{sec:LBmulti}. The points represent the steady velocity of migration obtained from the numerical simulations. To complete our benchmarking tests, the analysis reported in the right Panel of Fig.~\ref{fig:2} is extended to fluids with dissimilar thermal conductivities. Results are reported in Fig.~\ref{fig:3}. Overall, Figs.~\ref{fig:2}.-\ref{fig:3} reveal that the agreement between the measured velocity and the expected one is satisfactory, although the measured velocity tends to slightly underestimate the rhs of Eq.~\eqref{terminalbistxt} for the largest $\Ma$. This echoes the results of Liu {\it et al.} \cite{Zhangetal12}, in that the thermocapillary velocity normalized by the corresponding hydrodynamic prediction (valid at small $\Ma$) shows a decreasing behaviour in terms of $\Ma$, especially pronounced for $\Ma$ above a few units.\\ 
To conclude, we remark that in all the tests that we run, we noticed that a characteristic droplet diameter of the order of a few tens of grid points is enough to reproduce well the theoretical prediction of sharp interface hydrodynamics. Since 3D numerical simulations are affordable with such resolutions, this fact is instrumental to up-scale the problem and perform full 3D simulations in a confined T-junction in presence of thermocapillarity, as described in section \ref{sec:TJsetup}.

\section{Steady thermocapillary migration of a cylindrical droplet}\label{sec:Appendix}

In this appendix we provide the exact solution for the problem of a cylindrical droplet moving under the influence of thermocapillarity in a constant temperature gradient. Following~\cite{Subramanian01}, dealing with the case of a spherical drop, we employ a series of assumptions. First, we consider the system to be steady, and also assume that the non-linear effects on the momentum and heat transport are negligible, i.e. the Stokes flow and the Laplace equations hold for the velocity and temperature, respectively. The droplet then moves at a constant velocity $\vs$ in the direction ($x$) of the temperature gradient. The droplet is centered at the origin, and the temperature profile far away from the droplet is set by $T_\infty=T_0+x (\Delta T)/R$. We also assume a negligible deformation for the droplet and consider an unperturbed cylindrical shape. We then use a cylindrical coordinates system $(r,\theta)$ taken from the center of the droplet. The equations of continuity are expressed as
\begin{equation}
\frac{1}{r}\frac{\partial (r u_{r,i})}{\partial r}+\frac{1}{r}\frac{\partial u_{\theta,i}}{\partial\theta}= 0 \hspace{.2in} i={\rm c,d}.
\label{eq:b_cont01}
\end{equation}
The steady Stokes and Laplace equations for both the continuous and dispersed phases read 
\begin{equation}\label{eq:b_mom01}
-\frac{\partial P_i}{\partial r}+\eta_i \left\{\frac{\partial}{\partial r}\left(\frac{1}{r}\frac{\partial (r u_{r,i})}{\partial r}\right)+\frac{1}{r^2}\frac{\partial^2 u_{r,i}}{\partial\theta^2}-\frac{2}{r^2}\frac{\partial u_{\theta,i}}{\partial\theta}\right\}=0 \hspace{.2in} i={\rm c,d}
\end{equation}
\begin{equation}\label{eq:b_mom02}
-\frac{1}{r}\frac{\partial P_i}{\partial\theta}
+\eta_{i} \left\{ \frac{\partial}{\partial r}\left(\frac{1}{r}\frac{\partial (r u_{\theta,i})}{\partial r}\right)+\frac{1}{r^2}\frac{\partial^2 u_{\theta,i}}{\partial\theta^2} +\frac{2}{r^2}\frac{\partial u_{r,i}}{\partial\theta}\right\}=0 \hspace{.2in} i={\rm c,d}
\end{equation}
\begin{equation}\label{eq:b_lap01}
\kappa_i^{(T)}\left\{\frac{1}{r}\frac{\partial}{\partial r}\left(r\frac{\partial T_i}{\partial r}\right) +\frac{1}{r^2}\frac{\partial^2 T_i}{\partial\theta^2}
\right\} = 0 \hspace{.2in} i={\rm c,d}.
\end{equation}
At the large distances from the drop, we must have 
\begin{equation}\label{eq:BCinfinity}
u_{r,{\rm c}} \overset{r\rightarrow \infty}{\sim} -\vs \cos\theta, \hspace{.2in} u_{\theta,{\rm c}} \overset{r\rightarrow \infty}{\sim} \vs\sin\theta, \hspace{.2in}  P_c \overset{r\rightarrow \infty}{\sim} P_\infty, \hspace{.2in}   \frac{(T-T_0)R}{(\Delta T)} \overset{r\rightarrow \infty}{\sim} r\cos\theta, 
\end{equation}
where $P_{\infty}$ denotes the pressure far from the drop. At the drop surface $r=R$, we need to impose the condition of zero normal velocity and the continuity of the tangential velocity component and the tangential stress 
\begin{equation}
u_{r,{\rm c}}=u_{r,{\rm d}}=0, 
\end{equation}
\begin{equation}
u_{\theta,{\rm c}}=u_{\theta,{\rm d}},
\end{equation} 
\begin{equation}\label{eq:tangential}
\etac \left(r\frac{\partial (u_{\theta,{\rm c}}/r)}{\partial r}+ \frac{1}{r}\frac{\partial u_{r,{\rm c}}}{\partial \theta} \right)- \etad\left( r\frac{\partial (u_{\theta,{\rm d}}/r)}{\partial r}+ \frac{1}{r}\frac{\partial u_{r,{\rm d}}}{\partial \theta} \right)= \left(-\frac{{\rm d}\sigma}{{\rm d}T}\right) \frac{1}{r} \frac{\partial T_c}{\partial\theta}
\end{equation}
where $\sigma$ denotes the surface tension.  These three conditions are supplemented with the continuity of temperature and the continuity of the heat flux, expressed as 
\begin{equation}
T_c=T_d \label{eq:b_contt01},
\end{equation} 
\begin{equation}
\kappac^{(T)}\frac{\partial T_c}{\partial r}=\kappad^{(T)}\frac{\partial T_d}{\partial r}.
\end{equation}
We assume that the derivative of the surface tension $\sigma$ with respect to the temperature, $-{{\rm d}\sigma}/{{\rm d}T}$, is constant. Since there is no gravity acceleration, the drag force acting on the drop is zero, indicating that no stokeslet is involved in the system~\cite{Subramanian01} and the velocity vector field outside the drop is irrotational, and thus identified from  the governing Eqs.~\eqref{eq:b_cont01},~\eqref{eq:b_mom01} and~\eqref{eq:b_mom02} and the boundary conditions~\eqref{eq:BCinfinity}
\begin{equation}
u_{r,{\rm c}}=-\vs\left(1-\frac{R^2}{r^2}\right)\cos\theta, \hspace{.2in}    
u_{\theta,{\rm c}}=\vs\left(1+\frac{R^2}{r^2}\right)\sin\theta,  \hspace{.2in} 
P_c=P_\infty.
\label{eq:b_uout}
\end{equation}
By neglecting the solution with singularity at the origin, the velocity components in the dispersed phase inside the droplet satisfying the above boundary conditions are
\begin{equation}
u_{r,{\rm d}}=\vs\left(1-\frac{r^2}{R^2}\right)\cos\theta, \hspace{.2in}   
u_{\theta,{\rm d}}=\vs\left(-1+\frac{3r^2}{R^2}\right)\sin\theta, \hspace{.2in}  
P_d = P_\infty+\frac{\sigma_0}{R}-\frac{8\eta_d \vs r\cos\theta}{R^2} 
\label{eq:b_uin}
\end{equation}
where $\sigma_0$ denotes the surface tension at the temperature of $T=T_0$. The solutions to~\eqref{eq:b_lap01} which satisfy the temperature conditions are 
\begin{equation}\label{eq:b_t}
T_c=T_0+ (\Delta T)\left( \frac{r}{R}+\frac{(\kappac^{(T)}-\kappad^{(T)})R}{(\kappac^{(T)}+\kappad^{(T)})r} \right)\cos\theta, \hspace{.2in} 
T_d=T_0+ \frac{2(\Delta T)\kappac^{(T)} r}{(\kappac^{(T)}+\kappad^{(T)})R}\cos\theta.
\end{equation}
Finally, substituting~\eqref{eq:b_uout},~\eqref{eq:b_uin} and~\eqref{eq:b_t} into the tangential stress balance condition~\eqref{eq:tangential}, we obtain the translational speed of the drop
\begin{equation}\label{terminal}
\vs=\frac{\kappac^{(T)}(\Delta T)}{2(\etac+\etad)(\kappac^{(T)}+\kappad^{(T)})} \left(-\frac{{\rm d}\sigma}{{\rm d}T}\right).
\end{equation}
Looking at Eqs.~\eqref{eq:b_t} and~\eqref{terminal}, we can rewrite the steady velocity as
\be\label{terminalbis}
\vs=\frac{\Delta \sigma_{2R}}{8 (\etac+\etad)}
\ee
\be\label{eq:Deltasigma}
\Delta \sigma_{2R}=\Delta \sigma_{2R}(\kappac^{(T)},\kappad^{(T)})=\left(-\frac{{\rm d}\sigma}{{\rm d}T}\right)\frac{4 \kappac^{(T)} (\Delta T)}{(\kappac^{(T)}+\kappad^{(T)})}=\left(-\frac{{\rm d}\sigma}{{\rm d}T}\right)(T_{\mbox{\tiny{front}}}-T_{\mbox{\tiny{rear}}})
\ee
where $\Delta \sigma_{2R}$ represents the variation in the surface tension between the rear and the front of the moving droplet.

\bibliographystyle{spphys}       

\bibliography{sample}

\end{document}